\documentclass[prx, twocolumn, superscriptaddress, longbibliography]{revtex4-1}
\usepackage{bm, amsmath, amsfonts, amssymb, ascmac, braket}
\usepackage[dvipdfmx]{graphicx}
\usepackage{float}
\usepackage{color}

\newcommand{\s}{\sigma}

\newcommand{\LE}{\left}
\newcommand{\R}{\right}
\newcommand{\p}{\varphi}

\begin{document}

\title{Geometric Criterion for Solvability of Lattice Spin Systems}

\author{Masahiro Ogura}
	\affiliation{Yukawa Institute for Theoretical Physics, Kyoto University, Kyoto 606-8502, Japan}
\author{Yukihisa Imamura}
	\affiliation{Yukawa Institute for Theoretical Physics, Kyoto University, Kyoto 606-8502, Japan}
\author{Naruhiko Kameyama}
	\affiliation{Graduate School of Mathematics, Nagoya University, Nagoya 464-8602, Japan}
\author{Kazuhiko Minami}
	\affiliation{Graduate School of Mathematics, Nagoya University, Nagoya 464-8602, Japan}
\author{Masatoshi Sato}
	\affiliation{Yukawa Institute for Theoretical Physics, Kyoto University, Kyoto 606-8502, Japan}
\date{\today}

\begin{abstract} 
We present a simple criterion for solvability of lattice spin systems on the basis of the graph theory and the simplicial homology.
The lattice systems satisfy  algebras with graphical representations.
It is shown that the null spaces of adjacency matrices of the graphs provide conserved quantities of the systems.
Furthermore, when the graphs belong to a class of simplicial complexes,  
the Hamiltonians are found to be mapped to bilinear forms of Majorana fermions, from which the full spectra of the systems are obtained.    
In the latter situation, we find a relation between conserved quantities and the first homology group of the graph, and
the relation enables us to interpret the conserved quantities as flux excitations of the systems. 
The validity of our theory is confirmed in several known solvable spin systems including the 1d transverse-field Ising chain, the 2d Kitaev honeycomb model and the 3d diamond lattice model.
We also present new solvable models on a 1d tri-junction, 2d and 3d fractal lattices, and the 3d cubic lattice.
\end{abstract}

\maketitle

\section{Introduction}
 Exactly solvable models have been played important roles in the understanding of physics in strongly correlated systems.
In particular, exactly solvable lattice spin models have revealed many important phenomena. 
For instance, solving the 2d Ising model exactly, Onsager \cite{onsager1944crystal} showed the presence of ferromagnetic phase transition in spin systems for the first time, which is one of milestones in statistical physics.  
Since Onsager's work, other lattice spin models were solved exactly, such as the Potts model, the hard-hexagon model, and so on \cite{baxter2016exactly,wu1971ising,kadanoff1971some}.
More recently, exactly solvable models also have disclosed exotic quantum phases in strongly correlated systems, such as spin liquid phases with non-abelian anyon excitations \cite{kitaev2006anyons}.

Quantum solvable lattice spin models are classified into three types.
The first one has a Hamiltonian of which terms commute with each other, which 
includes the 2d Kitaev's toric code \cite{kitaev2006anyons,kitaev2009topological}, the X-cube model \cite{castelnovo2010quantum,nandkishore2019fractons}, and so on. 
The second one has special symmetries such as Lie groups or quantum groups.
This type includes the 1d Heisenberg model and the XXZ model \cite{pasquier1990common}.
Then, the last one can be transformed into free-fermion systems \cite{jordan1928pauli,nambu1995note,lieb1961,niemeijer1967some,katsura1962statistical,perk1975soluble,minami2016solvable,minami2017infinite,imamura,prosko2017simple,kaufman1949crystal2,perk1975soluble,perk1984finite,perk2017onsager,kitaev2006anyons,feng2007topological,chen2007exact,chen2008exact,kitaev2009topological,minami2019honeycomb}.
For instance, both the 1d XY model and the 1d transverse field Ising model can be converted into free-fermion systems by using the Jordan-Wigner transformation.
Another example is the Kitaev's honeycomb lattice model, which 
is transformed into a free fermion system by adapting a redundant representation of spins with Majorana operators.

In this paper, we present a simple criterion for the third type of solvability of lattice spin systems.
Our criterion is based on the graph theory and the simplicial homology. 
For a lattice spin system with an algebra with a graphical representation, 
we show that the null space of the adjacency matrix of the graph provides conserved quantities of the system.
Furthermore, when the graph belongs to a class of simplicial complexes,  we reveal that
the Hamiltonian is mapped to a bilinear form of Majorana fermions, from which the full spectrum of the system is obtained.   
We also find a relation between the conserved quantities and the first homology group of the graph.
Based on the relation, we interpret the conserved quantities as flux excitations.
We apply our criterion for several known solvable spin systems including the 1d transverse-field Ising chain, the 1d XY model, the 2d Kitaev honeycomb model, and the 3d diamond lattice model. 
We also present new solvable models on a 1d tri-junction, 2d and 3d fractal lattices, and the 3d cubic lattice.


The rest of this paper is organized as follows.
In Sec.~\ref{sec:main}, we present the main results.
We introduce lattice models which satisfy a class of algebras.
Representing the algebra in the form of a graph,  we present Theorems that give the criterion of solvability in terms of the graph theory and the simplicial homology. 
In Sec.~\ref{sec:application}, we illustrate our criterion by applying it to the 1d transverse-field Ising model, the XY model, the Kitaev honeycomb model and so on.
We also provide new solvable models in Sec.\ref{sec:new}
In Sec.~\ref{sec:proof}, we present proofs of Theorems in Sec.\ref{sec:main}.
We finally give discussions in Sec.\ref{sec:conclusion}.

%

\section{Main Results}
\label{sec:main}

First, we present our main results in this paper, which are summarized in three Theorems.
The proofs of these Theorems will be given in Sec.\ref{sec:proof}.

In this paper, we consider a class of Hamiltonians $H$ that satisfy the following properties.
\begin{itemize}
    \item $H$ has the form of $H=\sum_{j=1} ^n \lambda_j h_j$ with coefficients $\lambda_j\in \mathbb{R}$ and operators $h_j$ $(j=1,\dots, n)$.
\item The operators $h_j$ obey
$h_j ^2=1$,  
$h_j ^\dagger=h_j$, and $h_jh_k=\epsilon _{jk}h_kh_j$
with $\epsilon_{ij}=\pm 1$.
\end{itemize}
The second property requires that $h_j$s commute or anti-commute with each other. 
The operators $h_j$ generate an algebra ${\cal A}$ on $\mathbb{C}$, which we call the bond algebra (BA) \cite{nussinov2009bond,cobanera2011bond}. 
To represent the BA ${\cal A}$ visually, we introduce a graph ${\cal G(A)}$ as follows.
\begin{itemize}
    \item Put $n$ vertices in general position and place $h_i$ on the $i$-th vertex.
    \item When $h_i$ and $h_j$ anti-commute (commute) with each other,  we draw (do not draw) a line between the vertices with $h_j$ and $h_k$.
\end{itemize}
The resulting graph compactly encodes the information of the commutativity among $h_j$s.
We call the graph ${\cal G(A)}$ 
as commutativity graph (CG) \cite{wang2019tightening} of $\cal{A}$.
The CG ${\cal G(A)}$ has an algebraic representation with an adjacency matrix $M({\cal A})$. 
The adjacency matrix $M({\cal A})$
is a real symmetric $n\times n$ matrix of which elements indicate whether pairs of vertices are adjacent or not in ${\cal G(A)}$: The diagonal elements of $M({\cal A})$ are zero and the $(i,j)$-component is chosen to be $1$ ($0$) if $i$- and $j$-th vertices in ${\cal G(A)}$ are connected (not connected) by a line.
The multiplication and the addition for $M({\cal A})$ are defined as a matrix on the binary field $\mathbb{F}_2$, 
i.e. a matrix with entries $0$ or $1$, 
which satisfy $0+0=0$, $0+1=1$, $1+0=1$, and $1+1=0$.

Using $M({\cal A})$, we present our first main result.
A product $h_{j_1}h_{j_2}\cdots h_{j_k}$ conserves if it commutes with any $h_j$ in $H$.  We find that such conserved quantities in ${\cal A}$ can be counted by using the adjacency matrix $M({\cal A})$. 
More precisely, we have Theorem 1:
\medskip
\begin{itembox}[l]{Theorem 1}
Let $\cal{A}$ be the BA of a Hamiltonian $H=\sum_{j=1} ^n \lambda_j h_j$,  $\cal{G(A)}$ be the corresponding CG of $\cal{A}$, and $M({\cal A})$ be the adjacency matrix of ${\cal G(A)}$.
Then, the dimension  of the kernel space of $M({\cal A})$
coincides with the total number of conserved quantities in the form of $h_{j_1}\cdots h_{j_k}$.   
\end{itembox}
\medskip

\noindent
Here the kernel space (or null space) of $M({\cal A})$ is defined by
\begin{align}
{\rm Ker}M({\cal A})= \{{\bm v}\in \mathbb{F}_2^n; 
M({\cal A}){\bm v}={\bm 0}\}. 
\end{align}
As is shown in Sec.\ref{sec:proof}, we can construct the conserved quantities from an element ${\bm v}$ of ${\rm Ker}M({\cal A})$: 
Let ${\bm v}(h_j)$ be the unit vector on $\mathbb{F}_2$ having a nonzero element only in the $j$-th component, 
\begin{align}
{\bm v}(h_j)=
\begin{pmatrix}
0 & \cdots & 0 &1&0\cdots& 0 \\
\end{pmatrix}^T.
\end{align}
We can uniquely decompose ${\bm v}\in {\rm Ker}M({\cal A})$
in the form of
\begin{align}
{\bm v}={\bm v}(h_{l_1})+{\bm v}(h_{l_2})+\cdots+{\bm v}(h_{l_m}).        
\end{align}
Then, $h_{l_1}h_{l_2}\cdots h_{l_m}$ is a conserved quantity of $H$.

\begin{figure}[tbp]
 \begin{center}
  \includegraphics[width=5cm]{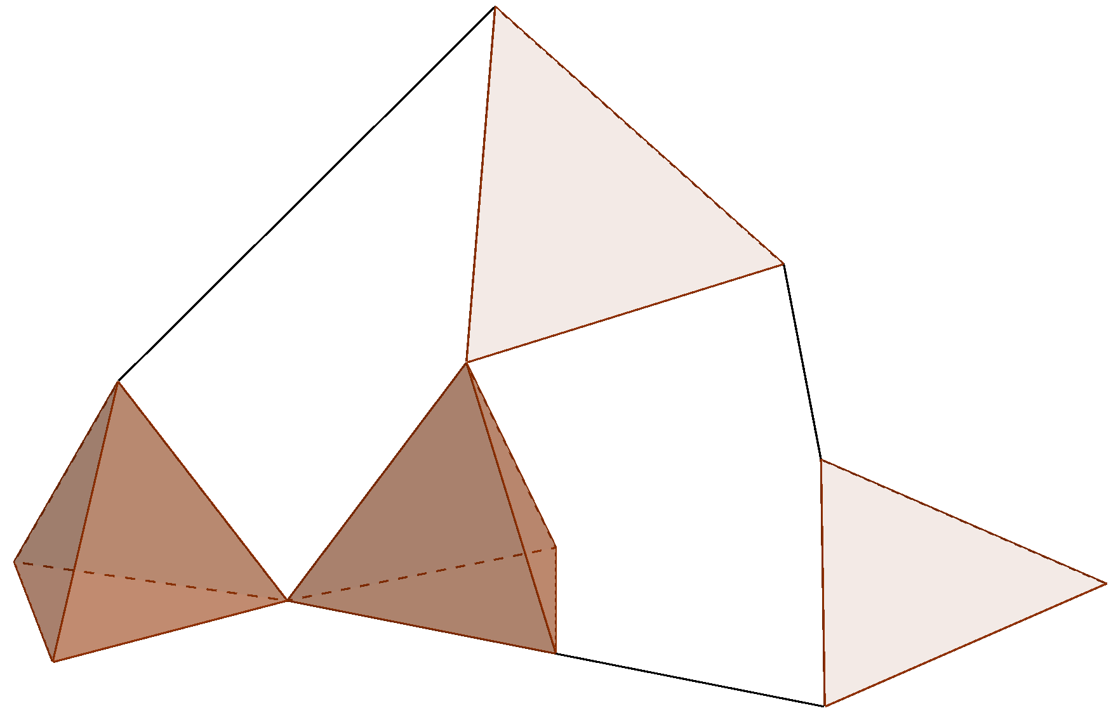}
 \end{center}
     \caption{A single-point-connected simplicial complex. Two 3-simplices (dark brown tetrahedrons) , Two 2-simplices (light brown triangles), and three 1-simplices (black lines) are connected only by vertices.}
 \label{fig:SPSC}
\end{figure}

The CG also enables us to characterize the BA geometrically.
For this purpose, we adapt the notion of simplex: 
A $d$-simplex is a $d$-dimensional polyhedron having the minimal number of vertices, namely $d+1$ vertices.
For instance, a 0-simplex is a vertex, a 1-simplex is a line, a 2-simplex is a triangle, a 3-simplex is a tetrahedron, and so on.
In particular, we consider a special set of simplices, which we call point-connected simplices:
Let us consider a set of simplices $S=\{s_1, \dots, s_m\}$ and let $V$ be a set consisting of all vertices of $s_\alpha\in S$ ($\alpha=1,\dots,m$).
Then, we call $S$ as point-connected if $S$ is connected and any pair of $s_\alpha, s_\beta\in S$ ($\alpha\neq \beta$) having a non-empty intersection shares only a single vertex $v\in V$ (Namely
$s_\alpha\cap s_\beta=\{v\}$).
Furthermore, we call $S$ as single-point-connected if any vertex $v\in V$ is shared by at most two different $s_\alpha$s.
Adding all faces of $s_\alpha\in S$ $(\alpha=1,\dots,m)$ into $S$, we obtain
a simplicial complex $K(S)$, which we dub single-point-connected simplicial complex (SPSC).
See Fig. \ref{fig:SPSC}.
Now we describe Theorem 2.

\medskip
\noindent
\begin{itembox}[l]{Theorem 2}
Let $\cal{A}$ be the BA of a Hamiltonian $H=\sum_{j=1} ^n \lambda_j h_j$ and $\cal{G(A)}$ be the corresponding CG of $\cal{A}$.
If ${\cal G(A)}$ coincides with a SPSC $K(S)$ with $S=\{s_1,\dots, s_m\}$, then $H$ is written by a bilinear form of $m$ Majorana operators. 
In particular, $h_j$ is recast into
\begin{eqnarray}
h_j=-i\epsilon_{\alpha\beta} \p_\alpha \p_\beta, \quad \epsilon_{\alpha\beta}=\pm 1,
\label{eq:hpp}
\end{eqnarray}
where $\varphi_{\alpha}$ are Majorana operators with the hermiticity 
$\p_\alpha ^\dagger=\p_\alpha$ 
and the anti-commutation relation 
$\{\varphi_{\alpha},\varphi_\beta\}=2\delta_{\alpha, \beta}$.
\end{itembox}
\medskip

Remarks are in order. (i) Without loss of generality, we can assume that any vertex $v$ of $s_{\alpha}\in S$ is shared by another $s_{\beta}\in S$ ($\beta\neq \alpha$): If not, 
we can add $v$ itself into $S$ as a 0-simplex to meet the assumption.
(ii) Under this assumption, the Majorana operator $\varphi_{\alpha}$ in Theorem 2 can be assigned to the simplex $s_{\alpha}\in S$. 
Then, $\varphi_{\alpha}$ and $\varphi_{\beta}$ in Eq.(\ref{eq:hpp}) are given by those on the simplices that share the vertex with $h_j$.
%
%
%
(iii)
The sign factors $\epsilon_{\alpha\beta}$ in Eq.(\ref{eq:hpp}) are determined as follows.
First, we use a sign ambiguity in Majorana operators:
We can multiply $\varphi_{\alpha}$ by $-1$ 
without changing the (anti-)commutation relations between them.
Using this gauge transformation, we can change the $m-1$ relative signs between $\varphi_{\alpha}$, which enables us to erase $m-1$ $\epsilon_{\alpha\beta}$s. 
There still, however, remain $n-m+1$ $\epsilon_{\alpha\beta}$s. 
The following Theorem 3 tells us that these remaining sign factors are determined by conserved quantities.
%
%

\medskip
\begin{itembox}[l]{Theorem 3}
Let ${\cal A}$ be the BA obeying the same assumption of Theorem 2. Then, 
$K(S)$ has independent $n-m+1$ non-contractible loops as a simplicial complex on $\mathbb{F}_2$.
Correspondingly, there exist $n-m+1$ conserved quantities that determine the remaining $n-m+1$ sign factors.
\end{itembox}
\medskip

It should be noted here that for each non-contractible loop, there remains        a sign factor that cannot be removed by the gauge transformation.
To count the number of independent non-contractible loops in $K(S)$, 
we calculate the homology group $H_q(K(S))$ of $K(S)$.
As we shall show in Sec.\ref{sec:proof}, a straightforward calculation shows that  $H_{q\ge 2}(K(S))=0$ and ${\rm dim}H_1(K(S))=n-m+1$ when $K(S)$ is a SPSC.
The latter result implies that $K(S)$ has $n-m+1$ independent non-contractible loops. 
We also find that each loop gives a conserved quantity: 
Take non-contractible loops as small as possible, then the product of all $h_j$s on each loop gives a conserved quantity.
Furthermore, we find that the conserved quantity reduces to the sign factor on the loop by rewriting it in terms of Majorana fermions in Eq.(\ref{eq:hpp}).

Theorems 2 and 3 imply that $H$ is solvable as a free Majorana system:
We can obtain the full spectrum of $H$ just by diagonalizing the free Majorana Hamiltonian.

We summarize the relation between the original spin model, the CG, the SPSC, and the free-fermion representation in Table \ref{table}.

%

\begin{table*}[tb]
\caption{Relations between the original model, the commutativity graph (CG), the single-point-connected simplicial complex (SPSC), and the free-fermion representation.}
\begin{tabular}{ccccccc}\hline \hline
original model&$\Leftrightarrow$ &CG $M({\cal A})$ &$\supset$ & SPSC $K(S)$ &$\Leftrightarrow$& free-fermion rep.   \\ \hline
$h_j$ &$\Leftrightarrow$ &vertex& &$v\in s_\alpha\cap s_\beta$
&$\Leftrightarrow$& $-i\epsilon_{\alpha\beta}\varphi_\alpha\varphi_\beta$\\ 
$\{h_i, h_j\}=0$ &$\Leftrightarrow$& line && -- && --\\
-- &&clique &$\Leftrightarrow$& $s_\alpha\in K(S)$&$\Leftrightarrow$& Majorana op. $\varphi_\alpha$\\
$[h, H]=0$ &$\Leftrightarrow$& ${\rm Ker}M({\cal A})$ &$\supset$ & $H_1(K(S))$ &$\Leftrightarrow$& flux $\epsilon$ \\ \hline \hline
\end{tabular}
\label{table}
\end{table*}




\section{Applications to known solvable models}
\label{sec:application}
In this section, we apply our theory to known solvable models, which confirms the validity of our criterion.
There are also a lot of solvable lattice models by our method. 
For example, we have checked our method in models in Refs.  \cite{minami2019honeycomb,minami2017infinite,nussinov2009bond,shi2009topological,imamura,prosko2017simple,lee2007edge,yu2008exactly,chen2019bosonization}.



\subsection{Transverse-Field Ising Model and Related Models}

First, we examine a class of spin models obeying the following BA with $n=2N$
\begin{align}
h_j ^2=1, \quad \{h_j,h_{j+1}\}=0, 
\nonumber\\
\LE[h_j,h_k\R]=0,\quad (j\neq k\pm 1).
\label{eq:TFIMclass}
\end{align}

\begin{figure}[htbp]
 \begin{center}
  \includegraphics[width=4cm]{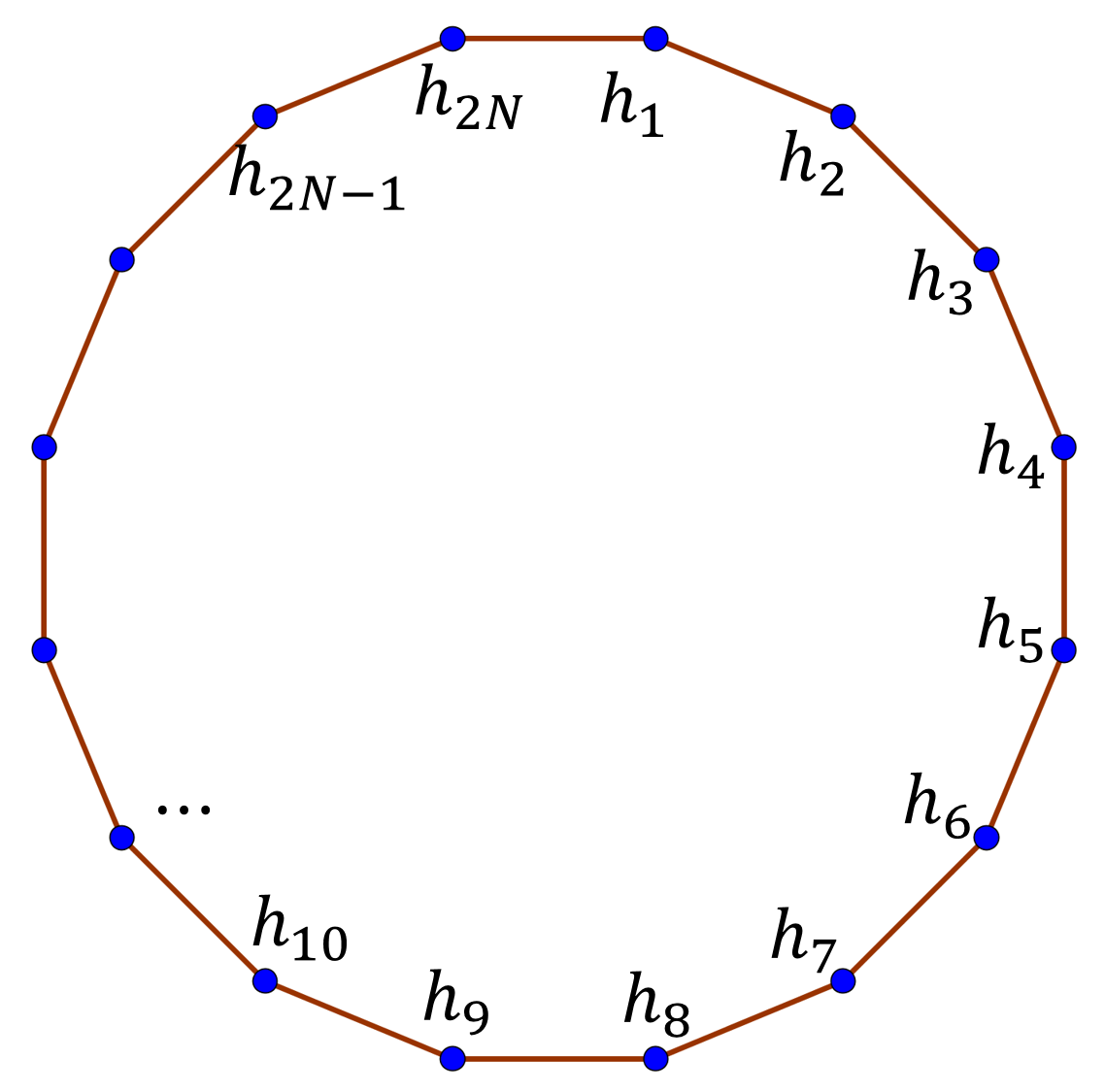}
 \end{center}
 \caption{The CG of Eq.(\ref{eq:TFIMclass}). The periodic boundary condition $h_{2N+1}=h_1$ is imposed.}
 \label{TFIM}
\end{figure}

In the periodic boundary condition $h_{2N+1}=h_1$, 
the CG of this algebra is a circle in Fig.\ref{TFIM}.
The corresponding adjacency matrix is given by
\begin{align}
M({\cal A})=
\begin{pmatrix}
0 & 1 & & & & 1\\
1& 0 & 1 &  &\text{{\huge{0}}} & \\
 &1 & \ddots & \ddots& & \\
  & &\ddots&\ddots&1&  \\
 &\text{\huge{0}}& &1 &0 &1\\
1& & &  &1&0
\end{pmatrix}.
\end{align}
For $N\ge 2$,
the kernel space of $M({\cal A})$ has the dimension $2$, which   
is spanned by $(1,0,1,0,\dots)^T$ and $(0,1,0,1,\dots)^T$.
Therefore, from Theorem 1,  we have two conserved quantities;
\begin{eqnarray}
c_1=h_1h_3\cdots h_{2N-1}, \quad
c_2=h_2h_4\cdots h_{2N}.
\label{eq:cTFIM}
\end{eqnarray}
Indeed, we can easily check that $c_1$ and $c_2$ commute with any $h_j$.
We also find that the CG in Fig. \ref{TFIM} is a SPSC.
Applying Theorem 2, we can rewrite $h_j$ in the form of  
\begin{eqnarray}
h_j=-i\epsilon_j \p_{j-1}\p_j,
\end{eqnarray}
where $\p_j$ is a Majorana operator and $\epsilon_j=\pm 1$.
Then, almost all $\epsilon_j$'s can be erased by redefining $\p_j$ as $\p_j\to \epsilon_j^{-1}\p_j$ ($j=1,\dots, 2N-1)$, and
after this, we obtain 
\begin{align}
&h_j=-i\p_{j-1}\p_j \quad (j=1,\cdots,2N-1),
\nonumber\\
&h_{2N}=-i\epsilon \p_{2N-1}\p_{2N}.   
\label{eq:TFIMclass2}
\end{align}
The remaining $\epsilon$ in Eq.(\ref{eq:TFIMclass2}) is determined by $c_1c_2$, 
\begin{eqnarray}
\epsilon=-c_1c_2.
\label{eq:eTFIM}
\end{eqnarray}
The sign factor $(-1)^N \epsilon$ corresponds to the $\pi$-flux through the hole of the CG in Fig.\ref{TFIM} \cite{kaufman1949crystal}.


In the open boundary condition, the CG is a line, and $M({\cal A})$ becomes
\begin{eqnarray}
M({\cal A})
=
\left(
\begin{array}{cccccc}
0 & 1 & 0 & 0 & \cdots & \\
1 & 0 & 1 & 0 & \cdots & \\
0 & 1 & 0 & 1 & \cdots & \\
0 & 0 & 1 & 0 & \cdots & \\
\vdots & \vdots & \vdots & \vdots & \ddots & \\
\end{array}
\right),
\end{eqnarray}
of which kernel is dimension 0 for $n=2N$. 
Now no conserved quantity is obtained, and thus  $\epsilon=1$.
In particular, in this case, our method naturally reproduces the Jordan-Wigner transformation \cite{minami2016solvable}.  
We can transform $M({\cal A})$ into the following form 
\begin{align}
Q^T M({\cal A})Q
=
\left(
\begin{array}{cccccc}
0 & 1 & 1 & 1 & \cdots & \\
1 & 0 & 1 & 1 & \cdots & \\
1 & 1 & 0 & 1 & \cdots & \\
1 & 1 & 1 & 0 & \cdots & \\
\vdots & \vdots & \vdots & \vdots & \ddots & \\
\end{array}
\right),
\end{align}
where $Q$ is given by
\begin{align}
Q
=\prod_{p}P^{[p,p+1]}=
\left(
\begin{array}{cccccc}
1 & 1 & 1 & 1 & \cdots & \\
0 & 1 & 1 & 1 & \cdots & \\
0 & 0 & 1 & 1 & \cdots & \\
0 & 0 & 0 & 1 & \cdots & \\
\vdots & \vdots & \vdots & \vdots & \ddots & \\
\end{array}
\right),
\end{align}
where $P^{[p,q]}$ is an elementary matrix with the $(i,j)$-component $P_{ij}^{[p,q]}=\delta_{ij}+\delta_{ip}\delta_{jq}$.
As we shall show in Sec.\ref{sec:proof}, $P^{[p,q]}$ induces a map  
\begin{align}
\{\ldots h_{p}, \ldots, h_{q}, \ldots\}
\mapsto 
\{\ldots h_{p}, \ldots, h_{p}h_{q}, \ldots\},    
\end{align}
and thus $Q$ gives a new bases 
\begin{eqnarray}
e_{j}=h_{1}h_{2}\cdots h_{j}.
\end{eqnarray}
The commutation relations in $Q^TM({\cal A})Q$ are 
$e_{i}e_{j}=-e_{j}e_{i}$ for all $i\neq j$,
those of the Clifford algebra. 
Introducing the initial operator $h_{0}$ that obeys $h_0^2=-1$, $\{h_0, h_1\}=0$ and $[h_0, h_j]=0$ ($j\neq 1$), and defining $\varphi_j$ as
\begin{eqnarray}
\varphi_{j}=i^{j-1}h_{0}h_{1}h_{2}\cdots h_{j}, 
\label{varphij}
\end{eqnarray}
we reproduces Eq.(\ref{eq:TFIMclass2}) with $\epsilon=1$.  
Equation (\ref{varphij}) is an algebraic generalization of the Jordan-Wigner transformation \cite{minami2016solvable}.
Actually, in the case of the transverse Ising chain below, 
by taking the initial operator as $h_{0}=i\sigma^{x}_{1}$, 
Eq.(\ref{varphij}) reproduces the original Jordan-Wigner transformation. 

For simplicity,  we only consider the periodic boundary condition below.



\subsubsection{Transverse-Field Ising Chain}
The Hamiltonian of the transverse-field Ising chain is given by
\begin{eqnarray}
H=-J\sum _{j=1} ^N \s _j ^x \s _{j+1} ^x -h\sum_{j=1} ^N \s _j ^z,
\label{eq:TFIM}
\end{eqnarray}
where $J$ is the exchange constant and $h$ is a transverse magnetic filed.
From Eq.(\ref{eq:TFIM}), the generator of the BA reads 
\begin{eqnarray}
h_{2j-1}=\s_j ^z, \quad h_{2j}=\s_{j} ^x\s_{j+1} ^x,
\label{hj-trIsing}
\end{eqnarray}
which satisfies Eq.(\ref{eq:TFIMclass}).
The conserved quantities in Eq.(\ref{eq:cTFIM}) are given by
\begin{align}
c_1=\prod_{j=1}^N \sigma_j^z. 
\quad c_2=1,
\end{align}
and thus the sign factor in Eq.(\ref{eq:eTFIM}) is
\begin{align}
\epsilon=-\prod_{j=1}^N \sigma_j^z.
\end{align}
From Eq.(\ref{eq:TFIMclass2}), the Hamiltonian is recast into
\begin{align}
H&=h\sum_{j=1}^N i \varphi_{2j-2}\varphi_{2j-1}
+J\sum_{j=1}^{N-1}i\varphi_{2j-1}\varphi_{2j}
\nonumber\\
&+Ji\epsilon \varphi_{2N-1}
\varphi_{2N},   
\label{eq:TFIMhamiltonian}
\end{align}
which reproduces the result in Ref.\cite{minami2016solvable}.

\subsubsection{Orbital Compass Chain}
\label{sec:occ}

Another model obeying Eq.(\ref{eq:TFIMclass}) is the orbital compass chain,
\begin{eqnarray}
H=-J_x\sum _{j=1} ^N \s _{2j-1} ^x \s _{2j} ^x -J_y\sum_{j=1} ^N \s _{2j} ^y\s_{2j+1} ^y,
\label{eq:OCCM}
\end{eqnarray}
where Eq.(\ref{eq:TFIMclass}) is obtained by the following identification,
\begin{align}
h_{2j-1}=\sigma_{2j-1}^x\sigma_{2j}^x,
\quad
h_{2j}=\sigma_{2j}^y\sigma_{2j+1}^y.
\end{align}
The conserved quantities $c_1$ and $c_2$ in Eq.(\ref{eq:cTFIM}) 
become
\begin{align}
c_1=\prod_{j=1}^{2N}\sigma_j^x,
\quad
c_2=\prod_{j=1}^{2N}\sigma_j^y,
\label{eq:cOCCM}
\end{align}
and thus $\epsilon$ in Eq.(\ref{eq:eTFIM}) is 
\begin{align}
\epsilon=(-1)^{N+1}\prod_{j=1}^{2N}\sigma_j^z.    
\end{align}
In terms of Majorana operators, $H$ in Eq.(\ref{eq:OCCM}) is given by
\begin{align}
H=&J_x\sum_{j=1}^N i\varphi_{2j-2}\varphi_{2j-1}
+J_y\sum_{j=1}^{N-1}i\varphi_{2j-1}\varphi_{2j}
\nonumber\\
&+J_yi\epsilon \varphi_{2N-1}\varphi_{2N},     
\end{align}
which coincides with Eq.(\ref{eq:TFIMhamiltonian}) if we identify $J_x$ and $J_y$ with $h$ and $J$.
Therefore, there is a one-to-one correspondence between the spectrum of the orbital compass chain and that of the transverse-field Ising chain.

On the other hand, there exist additional degeneracies in the orbital compass chain.
First,  $c_2$ in Eq.(\ref{eq:cOCCM}) can be $\pm 1$, which gives two-fold degeneracy of each state.
Moreover, we also have additional $2^N$-fold degeneracy.
This originates from the mismatch between the original spin degrees of freedom and the transformed Majorana degrees of freedom:
The original spin space is $2^{2N}$-dimensional, while the space of Majorana fermions is $2^N$-dimensional.
Correspondingly, 
there are additional conserved quantities $d_j$ ($j=1,\dots,2N$) which cannot be written by $h_j$,
\begin{eqnarray}
d_{2j-1}=\sigma_{2j-1}^y\sigma_{2j}^y,
\quad
d_{2j}=\sigma_{2j}^x\sigma_{2j+1}^x.
\end{eqnarray}
They satisfy the same BA as $h_j$;
\begin{eqnarray}
d_j ^2=1, \quad \{d_j,d_{j+1}\}=0, 
\nonumber\\
\LE[d_j,d_k\R]=0,\quad (j\neq k\pm 1), 
\end{eqnarray}
and thus these operators are equivalent to $2N$ Majorana fermions. As a result, they generate additional $2^N$-fold degeneracy.




\subsection{XY Model and Related Models}

\begin{figure}[htbp]
 \begin{center}
  \includegraphics[width=70mm]{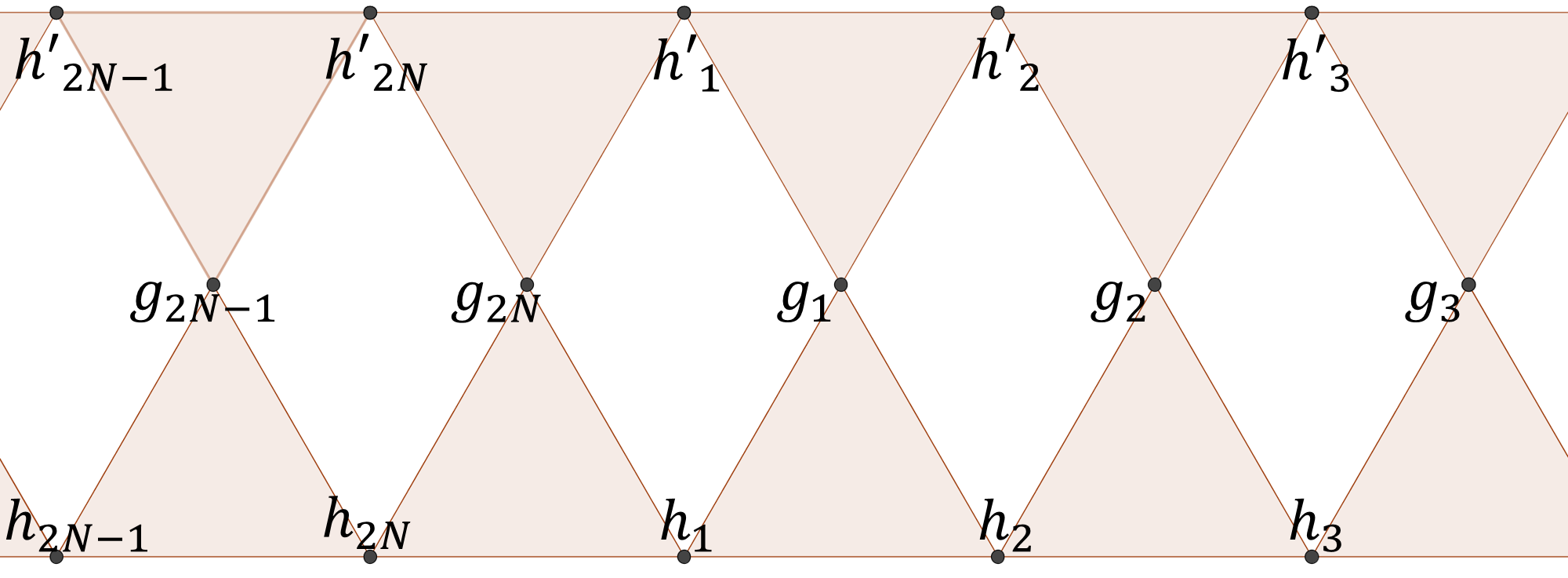}
 \end{center}
 \caption{The CG of Eq.(\ref{eq:XYclass})}
 \label{XY}
\end{figure}

Let $h_j$, $h_j'$, and $g_j$ ($j=1,\dots, 2N$)   
be operators obeying
\begin{align}
&h_j ^2=(h_j') ^2 =g_j^2=1,
\quad
\{h_{j},h_{j+1}\}=\{h_{j}', h_{j+1}'\}=0,
\nonumber\\
&\{h_{j},g_{j}\}=\{h_{j}',g_{j}\}=
\{h_{j+1},g_{j}\}=\{h_{j+1}',g_j\}=0,
\label{eq:XYclass}
\end{align}
where the other relations are commutative and the periodic boundary condition is assumed, 
\begin{align}
h_{i+2N}=h_i,\quad  h_{i+2N}'=h_i', \quad g_{i+2N}=g_i. 
\end{align}
This algebra defines a class of models with the CG in Fig. \ref{XY}.
The dimension of the kernel space of the adjacency matrix is $2N+2$, and we have $2N+2$ conservative quantities:
\begin{align}
&c_h=h_1\cdots h_{2N}, \quad
c_{h'}=h_1'\cdots h_{2N}', \quad
c_g=g_1\cdots g_{2N},    
\nonumber\\
&c_j=g_{j-1}h_j'g_jh_j \quad (j=1,\dots, 2N),
\label{eq:XYconserved}
\end{align}
which satisfy
\begin{eqnarray}
c_hc_{h'}c_1\cdots c_{2N}=1.
\end{eqnarray}
Since the CG in Fig.\ref{XY} is a SPSC, 
the operators in Eq.(\ref{eq:XYclass}) can be written by Majorana operators. 
Using the sign ambiguity (gauge degrees of freedom)  of Majorana operators,  we have  
\begin{align}
&h_j=-i\p_{j-1}\p_{j}, \quad h_j'=-i\p_{j-1}'\p_{j}'  
\nonumber\\
&g_j=-i\epsilon_j \p_j \p_j' \quad (j=1,\dots,2N-1), 
\nonumber\\
&h_{2N}=-i\epsilon \p_{2N-1}\p_{2N}, \quad
h_{2N}'=-i\epsilon' \p_{2N-1}'\p_{2N}',
\nonumber\\
&g_{2N}=-i\varphi_{2N}\varphi_{2N}',
\label{eq:varphiXY}
\end{align}
where $\p_i$ and $\p_i'$ are Majorana operators.
The sign factors $\epsilon_j$, $\epsilon$  and $\epsilon'$ are determined by the conserved quantities in Eq.(\ref{eq:XYconserved}),                       \begin{align}
\epsilon_j=\prod_{k=1}^{j}c_k, \quad 
\epsilon=(-1)^N c_h, \quad \epsilon'=(-1)^Nc_{h'}. 
\end{align}

\subsubsection{XY Model}
As a prime example of models with the CG in Fig.\ref{XY}, we consider the XY model,
\begin{equation}
H=-J\sum_{i=1} ^{2N} \left \{(1+\gamma)\sigma_i ^x \sigma_{i+1} ^x+(1-\gamma)\sigma_i ^y \sigma_{i+1} ^y\right \}-h\sum_{i=1} ^{2N} \sigma_i ^z,
\end{equation}
where $J$ is the exchange constant, $\gamma$ is the asymmetric parameter, and $h$ is a magnetic field. Actually, with the following identification 
\begin{align}
&h_{2j-1}=\sigma_{2j-1} ^x \sigma_{2j} ^x, \quad 
h_{2j}=\sigma_{2j} ^y \sigma_{2j+1} ^y, 
\nonumber\\
&h_{2j-1}'=\sigma_{2j-1} ^y \sigma_{2j} ^y, \quad
h_{2j}'=\sigma_{2j} ^x \sigma_{2j+1} ^x, 
\nonumber\\
&g_j=\s _{j+1}^z,    
\end{align}
we reproduce the BA in Eq.(\ref{eq:XYclass}).
In this model, the conserved quantities obey
\begin{align}
c_1=\cdots=c_{2N}=1, \quad
c_h=c_{h}'=-c_g=-\prod _{j=1} ^{2N} \s_j ^z,
\end{align}
and thus we have
\begin{align}
\epsilon_j=1, \quad \epsilon=\epsilon'
=(-1)^{N+1}\prod_{j=1}^{2N}\sigma_j^z.      
\end{align}
Therefore, Eq.(\ref{eq:varphiXY}) leads to
\begin{align}
H&=iJ\sum_{j=1} ^{N} \left \{(1+\gamma)(\p_{2j-2}\p_{2j-1}+\p_{2j-1}'\p_{2j}') \right \} \nonumber\\
&+iJ\sum_{j=1} ^N\left \{(1-\gamma)(\p_{2j-1}\p_{2j}+\p_{2j-2}'\p_{2j-1}') \right \}\nonumber\\
&+ih\sum_{j=1} ^{2N} \p_j\p_j' \nonumber \\
&-iJ(1-\epsilon)\left \{(1+\gamma) \p_{2N-1}\p_{2N} +(1-\gamma) \p_{2N-1}'\p_{2N}' \right \} .
\label{eq:XY2}
\end{align}
Equation (\ref{eq:XY2}) reproduces the known fermion representation of the XY model:
Introducing the fermion operators $a_j$ as
\begin{align}
\p_{2j-1}&=u_{2j-1}(a_{2j-1}+a_{2j-1} ^\dagger) , \nonumber\\
\p_{2j-1}'&=iu_{2j-1}(a_{2j-1}-a_{2j-1} ^\dagger) , \nonumber\\
\p_{2j}&=-iu_{2j}(a_{2j}-a_{2j} ^\dagger) , \nonumber\\
\p_{2j}'&=u_{2j}(a_{2j}+a_{2j} ^\dagger),
\end{align}
with $u_j=(-1)^{j(j-1)/2}$, 
we obtain 
\begin{align}
H=&-2J\sum_{j=1} ^{2N-1} \left[(a_j ^\dagger a_{j+1}+a_{j+1} ^\dagger a_j) 
+\gamma (a_j ^\dagger a_{j+1} ^\dagger +a_{j+1}a_j)   \right] \nonumber \\
&-2h\sum_{j=1} ^{2N} \left(a_j ^\dagger a_j-\frac{1}{2} \right) 
\nonumber\\
&+2Jc_g \left[(a_j ^\dagger a_{j+1}+a_{j+1} ^\dagger a_j)+\gamma (a_j ^\dagger a_{j+1} ^\dagger +a_{j+1}a_j )\right],
\end{align}
which is the same fermion reprentation in Ref. \cite{niemeijer1967some}.



\subsubsection{Ladder Model}
The second example is the ladder model \cite{degottardi2011topological},
\begin{align}
H=&-J_{\rm t}\sum_{j=1} ^{N}\LE(\s _{2j-1} ^x\s_{2j} ^x+
\s _{2j} ^y\s_{2j+1} ^y
\R)
\nonumber\\
&-J_{\rm b}\sum_{j=1} ^{N}\LE(\tau _{2j-1} ^x\tau_{2j} ^x
+\tau _{2j} ^y\tau_{2j+1} ^y
\R)
\nonumber\\
&-J_{\perp}\sum_{j=1} ^{2N} \LE(\s _j ^z \tau_j ^z\R),
\end{align}
where $J_{\rm t}$ ($J_{\rm b}$) is the intra exchange constant between top (bottom) spin chains, and $J_{\perp}$ is the inter exchange constant between top and bottom chains.  
This model gives 
\begin{align}
&h_{2j-1}=\s _{2j-1} ^x\s_{2j} ^x, \quad h_{2j}=\s _{2j} ^y\s_{2j+1} ^y, 
\nonumber\\
&h_{2j-1}'=\tau _{2j-1} ^x\tau_{2j} ^x, \quad
h_{2j}'=\tau _{2j} ^y\tau_{2j+1} ^y, 
\nonumber\\
&g_j=\s _j ^z \tau_j ^z,    
\end{align}
which satisfy Eq.(\ref{eq:XYclass}). 
In this model, we have
\begin{align}
&c_h=-\prod_{j=1} ^{2N} \s_j ^z, \quad 
c_{h'}=-\prod_{j=1} ^{2N} \tau_j ^z, \quad
c_g=c_hc_{h'},
\nonumber\\
&c_{2j-1}=-\s_{2j-1} ^y \s_{2j} ^y \tau_{2j-1} ^y \tau_{2j} ^y,
\nonumber\\
&c_{2j}=-\s_{2j} ^x\s_{2j+1} ^x \tau_{2j} ^x \tau _{2j+1} ^x, 
\end{align}
which lead to
\begin{align}
&\epsilon_{2j-1}=-\s_{1} ^y\tau_{1} ^y \left( \prod_{k=2} ^{2j-1} \s_k ^z \tau_k ^z \right) \s_{2j} ^y \tau_{2j} ^y , \quad \nonumber \\
&\epsilon_{2j}=-\s_1 ^y\tau_1 ^y \left( \prod_{k=2} ^{2j} \s_k ^z \tau_k ^z \right) \s_{2j+1} ^x \tau_{2j+1} ^x
\nonumber\\
&\epsilon'=(-1)^{N+1}\prod_{j=1}^{2N}\sigma_j^z, \quad\epsilon=(-1)^{N+1}\prod_{j=1}^{2N}\tau_j^z,
\end{align}
where $\prod_{k=2} ^{1} \s_k ^z \tau_k ^z \equiv 1$.
The Hamiltonian is equivalent to
\begin{align}
    H&=iJ_{\rm t}\sum_{j=1} ^{2N-1} \p_{j-1} \p_{j}+iJ_{\rm t}\epsilon \p_{2N-1} \p_{2N} 
    \nonumber\\
    &+iJ_{\rm b}\sum_{j=1} ^{2N-1} \p'_{j-1} \p'_{j} +iJ_{\rm b}\epsilon'\p_{2N-1}'\p_{2N}'
    \nonumber \\
    &+iJ_{\perp} \sum_{j=1} ^{2N-1} \epsilon_j\p_j\p_j'  
    +iJ_{\perp}\p_{2N}\p_{2N}'.  
\end{align}


\subsubsection{Double Spin-Majorana Model}

The third example is the double spin-Majorana model, 
\begin{align}
H=&-i g\sum _{j=1} ^{2N} \left(\gamma _j \s_j ^x\gamma _{j+1}+\gamma_j'\tau_j ^x \gamma_{j+1}'\right)
\nonumber\\
&-J\sum_{j=1} ^{2N}\s _j ^z \s _{j+1} ^z \tau_j ^z\tau_{j+1} ^z,
\end{align}
where $g$ and $J$ are real parameters, and $\gamma_j$'s are Majorana operators.　
The BA of this model reads
\begin{align}
&h_j=i\gamma _j \s_j ^x\gamma _{j+1}, \quad
h_j'=i\gamma_j'\tau_j ^x \gamma_{j+1}', 
\nonumber\\
&g_j=\s _j ^z \s _{j+1} ^z \tau_j ^z\tau_{j+1} ^z, 
\end{align}
which reproduces Eq.(\ref{eq:XYclass}), and 
we obtain 
\begin{align}
&c_h=(-1)^{N}\prod _{j=1} ^{2N} \s _j ^x, \quad
c_{h'}=(-1)^{N}\prod _{j=1} ^{2N} \tau _j ^x, \quad c_g=1,
\nonumber\\ 
&c_j=-\s _{j-1} ^z\tau _{j-1} ^z\s _j ^x\tau _j ^x \s _{j+1} ^z\tau _{j+1} ^z\gamma _j \gamma _{j}' \gamma_{j+1}  \gamma_{j+1}'. 
\end{align}
Therefore, 
\begin{align}
&\epsilon_1=-\sigma_{2N} ^z \tau_{2N} ^z \sigma_{1}^{x} \tau_1 ^x \s_2 ^z \tau_2 ^z \gamma_1 \gamma_1' \gamma_2 \gamma_2',
\nonumber\\
&\epsilon_j=-\sigma_{2N} ^z \tau_{2N} ^z \sigma_{1}^{y} \tau_1 ^y \left(\prod_{k=2} ^{j-1} \s_k ^z \tau_k ^z \right) 
\nonumber\\
&\times \s_j ^y \tau_j ^y \s_{j+1} ^z \tau_{j+1} ^z \gamma_1 \gamma_1' \gamma_{j+1} \gamma_{j+1}' \quad (j=2,\dots, 2N-1), 
\nonumber\\
\quad
&\epsilon=\prod_{j=1} ^{2N} \s_j ^z, \quad 
\epsilon'=\prod_{j=1} ^{2N} \tau_j ^z,
\end{align}
where $\prod_{k=2} ^1\s_k ^z\tau_k ^z\equiv 1$.
The Hamiltonian is recast into
\begin{align}
    H&=ig\sum_{j=1} ^{2N-1}\left(\p_{j-1}\p_{j}+\p_{j-1}'\p_{j}' \right) \nonumber \\
    &+ig\left(\epsilon\p_{2N-1}\p_{2N}+\epsilon' \p_{2N-1}'\p_{2N}' \right)
    \nonumber\\
    &+iJ\sum_{j=1} ^{2N-1} \epsilon_j \p_j \p_j' +iJ\p_{2N}\p'_{2N}.
\end{align}

In a manner similar to the orbital compass chain in Sec.\ref{sec:occ}, 
this model hosts additional degeneracies originating from the mismatch between the original degrees of freedom and the transformed Majorana ones:
It is found that the following operators $d_j$ and $d'_j$ ($j=1,\dots, 2N$) commute with $h_j,h_j',g_j$,
\begin{eqnarray}
d_j=\s_{j-1} ^z\gamma_j \s _j ^z, \quad
d_j'=\tau_{j-1} ^z\gamma_j' \tau _j ^z, 
\end{eqnarray}
which satisfies
\begin{eqnarray}
\{d_j,d_k\}=\{d_j',d_k'\}=2\delta_{j,k}, \quad 
\{d_j,d_k'\}=0.
\end{eqnarray}
Thus, each state of this model has $2^{2N}$-fold degeneracy.

\subsection{Kitaev Honeycomb Lattice Model}

\begin{figure}[htbp]
 \begin{center}
  \includegraphics[width=30mm]{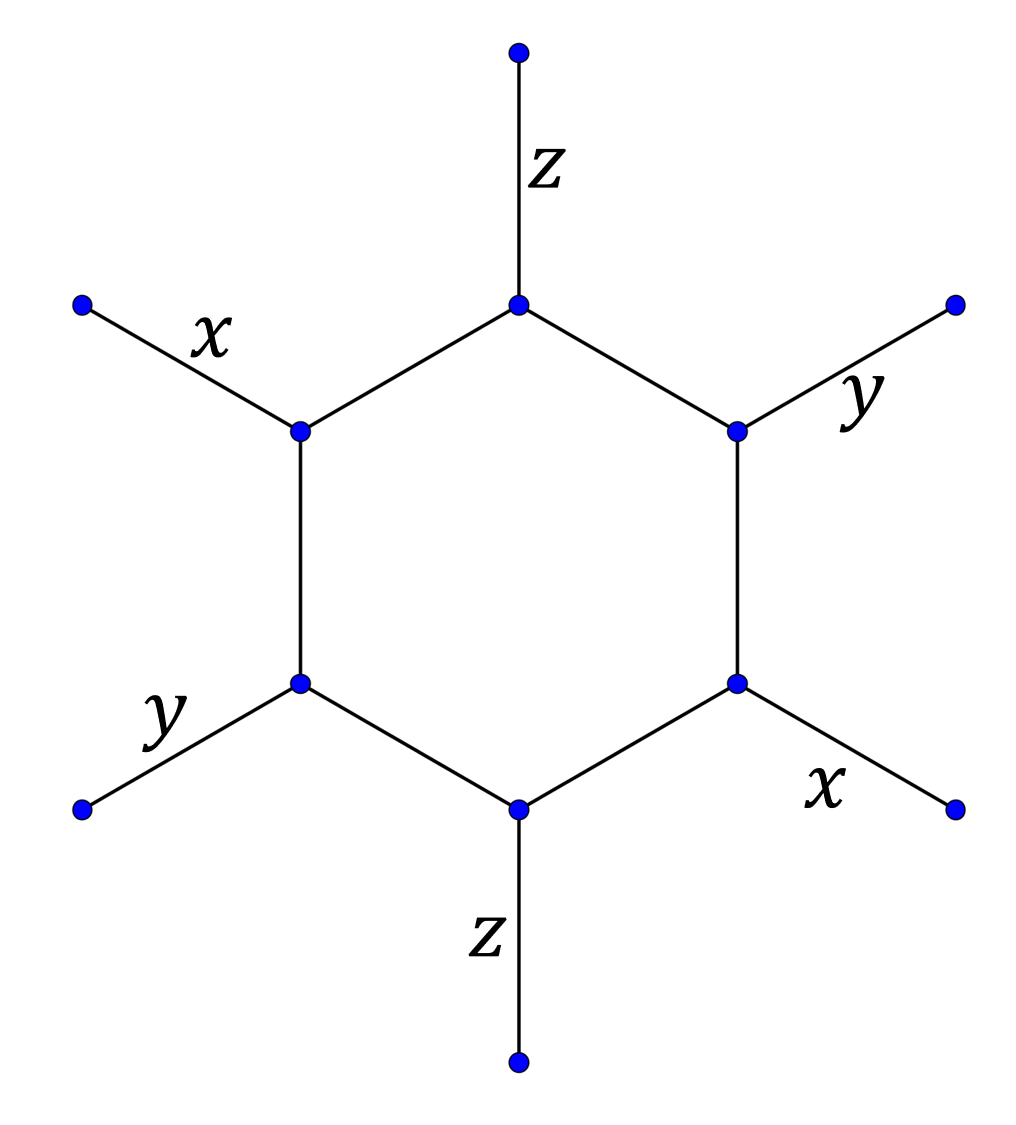}
 \end{center}
 \caption{$x$-, $y$- and $z$-links in honeycomb lattice.}
 \label{link}
\end{figure}

The Kitaev honeycomb lattice is described by the following Hamiltonian with the nearest neighbour spin couplings,
\begin{align}
H=&-J_x\sum _{\text{$x$-links}}\s _j ^x\s _k ^x
-J_y\sum _{\text{$y$-links}}\s _j ^y\s _k ^y
\nonumber\\
&-J_z\sum _{\text{$z$-links}}\s _j ^z\s _k ^z,
\label{eq:Kitaev}
\end{align}
where the orientation of the $x$, $y$, and $z$-links are indicated  in Fig.\ref{link}. 
\begin{figure}[htbp]
 \begin{center}
  \includegraphics[width=50mm]{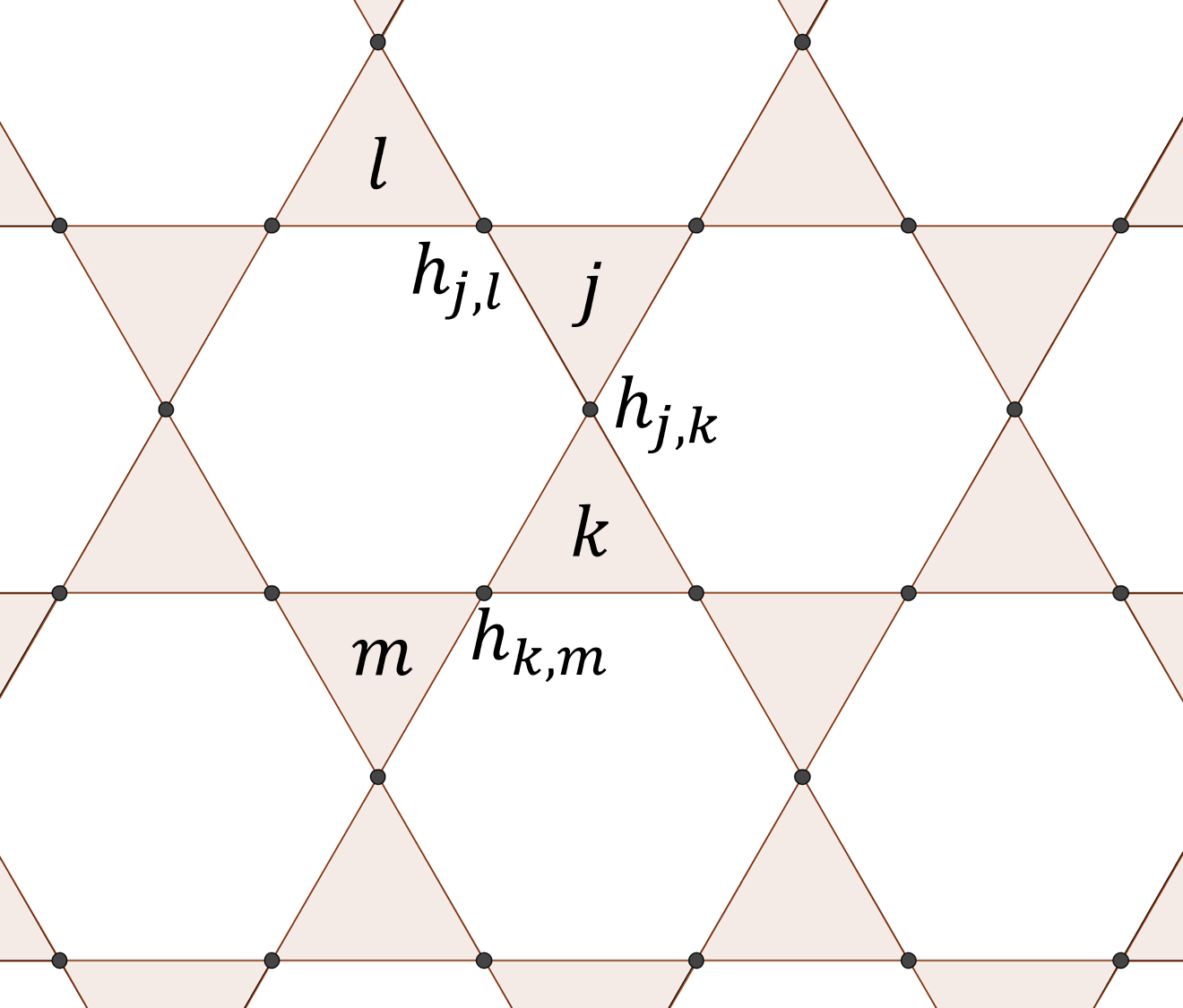}
 \end{center}
 \caption{The CG of the Kitaev honeycomb lattice model}
 \label{KHLM}
\end{figure}
Each term of Eq.(\ref{eq:Kitaev}) anti-commutes or commutes with each other, and thus it defines the BA. 
The CG of this model is the Kagome lattice in Fig. \ref{KHLM}.
The Kagome lattice is dual to the original honeycomb lattice, 
and each vertex in the Kagome lattice corresponds to a link in the honeycomb lattice. 
We assign an operator 
\begin{eqnarray}
h_{j,k}=\s_j ^{\mu(j,k)} \s_k ^{\mu(j,k)}
\end{eqnarray}
in the BA to each vertex of the Kagome lattice, where $\mu(j,k)=x,y,z$ is the spin-orientation at the corresponding $(j,k)$-link in the honeycomb lattice.
The conservative quantities are
\begin{align}
    c_p=\prod _{(j,k) \in \partial p} h_{j,k},
    \quad
    c_z=\prod _{(j,k): z-\text{link}} h_{j,k},
\label{eq:cp}
\end{align}
where $p$ is a hexagon in Fig. \ref{KHLM}

Regarding triangles in Fig. \ref{KHLM} as 2-simplices, 
the CG can be identified with a SPSC.
Therefore, we can apply Theorems 2 and 3 to the Kitaev honeycomb lattice model. The  operator $h_{j,k}$ is converted into a Majorana bi-linear form
\begin{align}
    h_{j,k}=-i\epsilon_{jk} \p_j \p_k,
\end{align}
so the Hamiltonian is equivalent to
\begin{eqnarray}
H=\sum_{\langle j, k\rangle}iJ_{\mu(j,k)}\epsilon_{jk}\p_j \p_k,
\end{eqnarray}
where $\epsilon_{ij}$'s are determined by the conserved quantities in Eq.(\ref{eq:cp}). 
This result reproduces that in Ref.\cite{kitaev2009topological}, although our derivation is much simpler than the original one.

\subsection{Diamond lattice model}
\begin{figure}[htbp]
 \begin{center}
  \includegraphics[width=40mm]{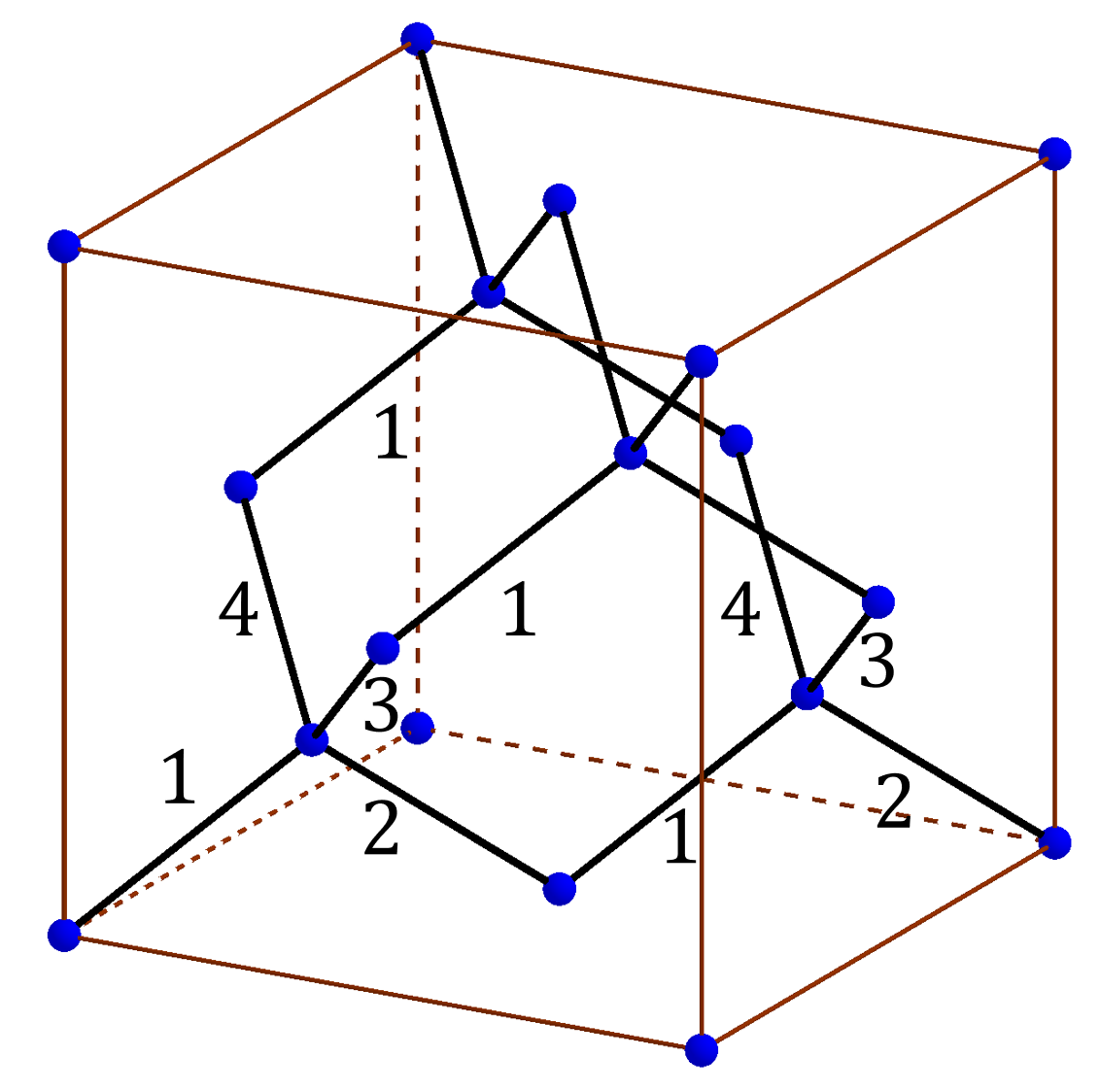}
 \end{center}
 \caption{Diamond lattice. The number at the link indicates the orientation $\mu$ of the gamma matrix in the diamond lattice model.}
 \label{diamond}
\end{figure}

\begin{figure}[htbp]
 \begin{center}
  \includegraphics[width=50mm]{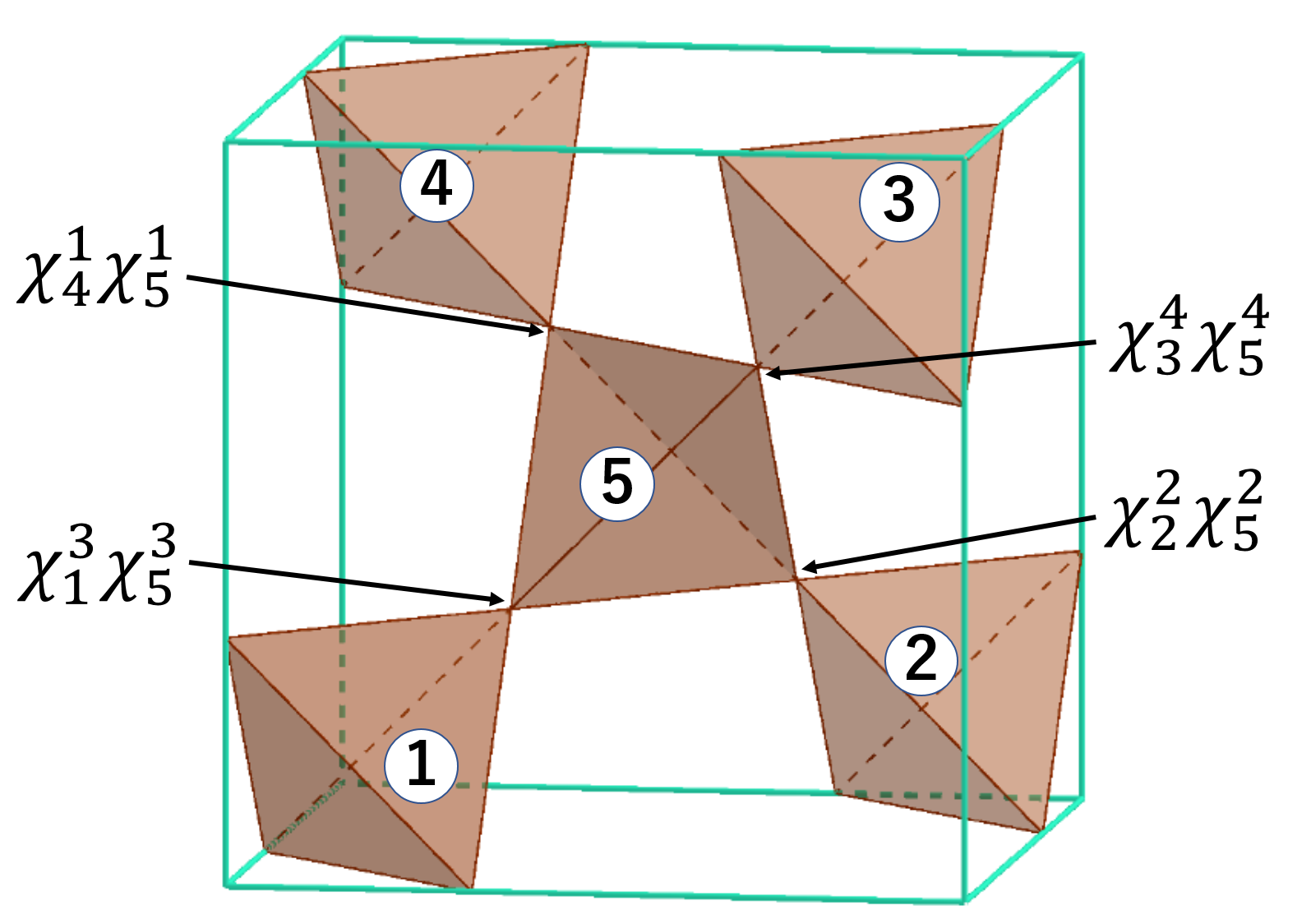}
 \end{center}
 \caption{The CG of the diamond lattice model with $\chi=\alpha,\zeta$.}
 \label{Pyrochlore}
\end{figure}

The diamond lattice is a three-dimensional analog of the honeycomb lattice \cite{ryu2009three,wu2009gamma}.
We can generalize the Kitaev honeycomb lattice model in three-dimensions.
The Hamiltonian is given by
\begin{eqnarray}
H=-\sum_{\langle j,k\rangle}J_{jk} \left( \alpha_{j} ^{\mu(j,k)}\alpha_{k} ^{\mu(j,k)}+\zeta_j ^{\mu(j,k)} \zeta_k ^{\mu(j,k)} \right),
\end{eqnarray}
where $\alpha_j^{\mu}$ and $\zeta_j^\mu$ $(\mu=1,2,3,4)$ are two sets of 
Dirac matrices, 
\begin{align}
&\alpha_j^a=\sigma_j^a\otimes \tau_j^x, \quad \alpha_j^4=\sigma_j^0\otimes \tau_j^z, 
\nonumber\\
&\zeta_j^a=-\sigma_j^a\otimes \tau_j^z, \quad \zeta_j^4=\sigma_j^0\otimes \tau_j^x,
\end{align}
with $a=1,2,3$, $j$ is the site index, and 
$\mu(j,k)=1,2,3,4$ indicates the orientation of the gamma matrix at $(j,k)$-link, as illustrated in Fig.\ref{diamond}.
We assign the operators $h_{j,k}$ and $h'_{j,k}$ as
\begin{align}
    h_{j,k}=\alpha_{j} ^{\mu(j,k)}\alpha_{k} ^{\mu(j,k)}, \quad
    h_{j,k}'=\zeta_j ^{\mu(j,k)} \zeta_k ^{\mu(j,k)},
\end{align}
which satisfy 
\begin{align}
\left[h_{j,k},h_{l,m}' \right]=0.
\end{align}
The CGs of $h_{j,k}$ and $h_{j,k}'$ are two identical pyrochlore lattices in Fig.\ref{Pyrochlore}.
By regarding tetrahedrons as 3-simplices, the pyrochlore lattice is identified with a SPSC.
From straightforward calculation, we also find that conserved quantities in two CGs are the same.
Therefore, we can transform $h_{j,k}$s and $h_{j,k}'$s into Majorana bi-linear forms, 
\begin{align}
    h_{j,k}=-i\epsilon_{j,k} \p_j \p_k,
    \quad
    h_{j,k}'=-i\epsilon_{j,k} \p_j' \p_k'.
\end{align}
Consequently, the Hamiltonian is converted into
\begin{align}
    H=i\sum_{\langle j,k\rangle}J_{jk}\epsilon_{j,k} \left(\p_j \p_k+ \p_j' \p_k' \right),
\end{align}
which reproduces that in \cite{ryu2009three}.



\section{New solvable models}
\label{sec:new}
So far, we have applied our method to known solvable models.
Our approach also provides a powerful method to construct new solvable models in variety of lattices.
In this section, we present such new solvable models.

\subsection{Tri-Junction Model}

\begin{figure}[htbp]
 \begin{center}
  \includegraphics[width=70mm]{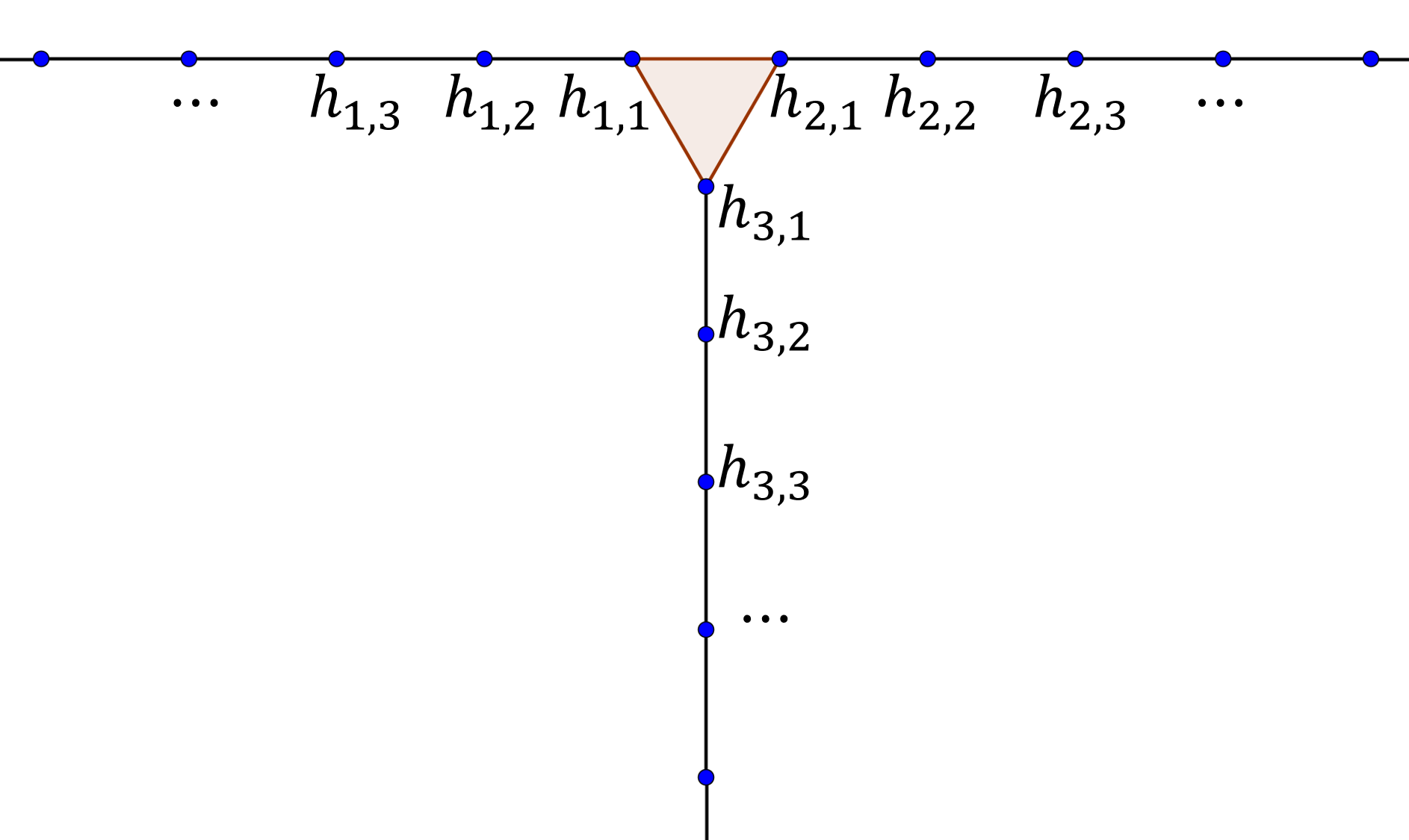}
 \end{center}
 \caption{The CG of the tri-junction model}
 \label{Junction}
\end{figure}

We first consider the transverse-field Ising chains with the tri-junction \cite{giuliano2016junction,backens2019jordan,giuliano2020emerging}.
The Hamiltonian is given by
\begin{eqnarray}
H=&-&\sum_{a=1} ^3\left[J_a\sum _{j=1} ^{N-1} \s _{a,j} ^z \s _{a,j+1} ^z +h_a\sum_{j=2} ^{N} \s _{a,j} ^x \right] \nonumber \\
&-&t_{12}\s_{1,1} ^x\s_{2,1} ^z -t_{23}\s_{2,1} ^x\s_{3,1} ^z -t_{31}\s_{3,1} ^x \s_{1,1} ^z,
\end{eqnarray}
where $J_a$ and $h_a$ are the exchange constant and a magnetic field of $a$-th chain, and $t_{ab}$ are the coupling between $a$-th and $b$-th chains.
The CG of this model is Fig. \ref{Junction}, where $h_{a,j}$ $(j=1,\dots,2N-1)$ is defined by
\begin{align}
    &h_{a,1}=\s_{a,1} ^x\s_{a+1,1} ^z,\nonumber\\
    &h_{a,2l}=\s_{a,l} ^z\s_{a,l+1} ^z,\quad h_{a,2l+1}=\s_{a,l+1} ^x.
\end{align}
From the adjacency matrix of the CG, we find a conserved quantity 
\begin{align}
    c&=-i\prod_{a=1} ^3 \prod_{j=1} ^{N} h_{a,2j-1} \nonumber \\
    &=\left(\prod_{a=1} ^3 \prod_{j=2} ^{N} \s_{a,j} ^x \right) \left(\prod_{a=1} ^3 \s_{a,1} ^y \right).
\end{align}

The CG in Fig.\ref{Junction} can be identified with a SPSC consisting of lines and a triangle.  
Therefore, applying Theorem 2 to this model, we have
\begin{align}
    &h_{a,1}=-i\p_{a,1} \p, \nonumber\\
    &h_{a,j}=-i\p_{a,j-1}\p_{a,j}, \quad (j=2,\dots, N).
\end{align}
By using this, the Hamiltonian is recast into the bilinear form of Majorana operators,
\begin{align}
    H&=i\sum_{a=1} ^3\left[J_a\sum _{j=1} ^{N} \p_{a,2j-1}\p_{a,2j} +ih_a\sum_{j=1} ^{N} \p_{a,2j}\p_{a,2j+1} \right] \nonumber \\
    &+\left(t_{12} \p_{1,1} +t_{23} \p_{2,1} +t_{31} \p_{3,1} \right) \p.
\end{align}
This model hosts implicit conserved quantities that is not obtained by $h_j$, 
\begin{align}
    c_a=\s_{a-1,1} ^z \prod_{j=1} ^N \s_{a,j} ^x \quad (a=1,2,3),
\end{align}
which satisfies
\begin{align}
    \left[c_a,h_{b,j} \right]=0, \quad
    \{c_a,c_b\}=2\delta_{a,b}, \quad
    ic_1c_2c_3=c.
\end{align}
These operators induce additional $2$-fold degeneracy.

By same method, we can construct n-junction model whose junction is a $n-1$-simplex.
We can also design tree-like models by junctions.

\subsection{Hanoi graph  model}
\begin{figure}[tbp]
 \begin{center}
  \includegraphics[width=45mm]{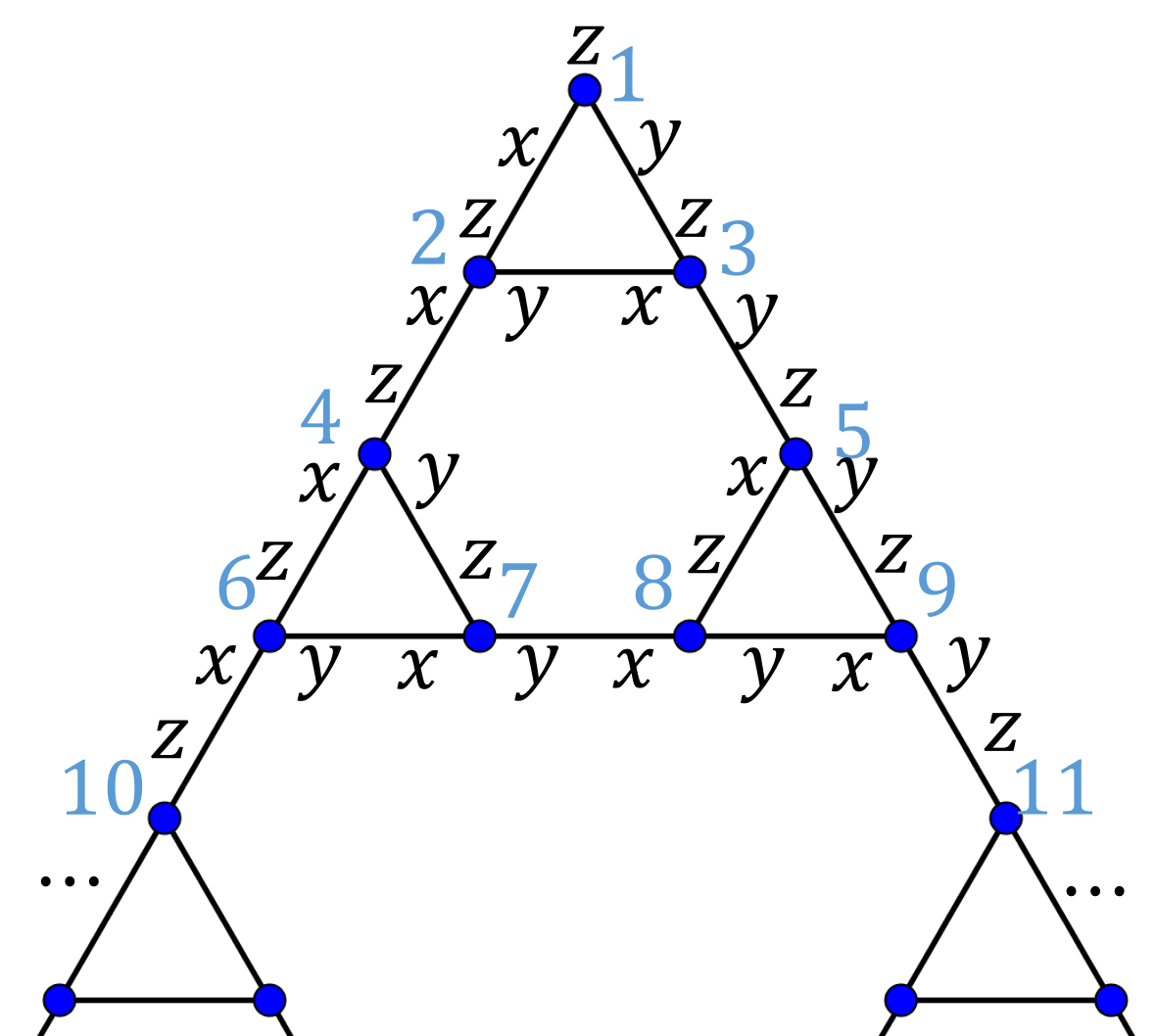}
 \end{center}
 \caption{Hanoi graph. $x$, $y$, and $z$ on each site denote the spin-orientation of the exchange interaction.}
 \label{Hanoi}
\end{figure}

We can construct solvable models in 2d and 3d fractal lattices.
Let us consider the Hanoi graph in Fig.\ref{Hanoi}, 
and place a spin operator on each site of the Hanoi graph.
Then, we consider the Hamiltonian
\begin{align}
H=&-J_1\s_1 ^z \nonumber\\
&-J_{12}\s_1 ^x\s_2 ^z-J_{13}\s_1 ^y\s_3 ^z \nonumber\\
&-J_{23}\s_2 ^y\s_3 ^x-J_{24}\s_2 ^x\s_4 ^z -J_{35}\s_3 ^y\s_5 ^z 
\nonumber\\
&-\cdots,    
\label{eq:Hanoi}
\end{align}
where $\sigma^\mu_i$ is the $\mu$-th Pauli matrix at the $i$-th site in Fig.\ref{Hanoi}, and $J_{ij}$ is the exchange constant.
The spin-orientation of the exchange interaction is determined as illustrated in Fig.\ref{Hanoi}: 
In the case of the (1,2) link, for instance, we take $\sigma^x$ and $\sigma^z$ from site 1 and site 2, respectively.  

\begin{figure}[htbp]
 \begin{center}
  \includegraphics[width=45mm]{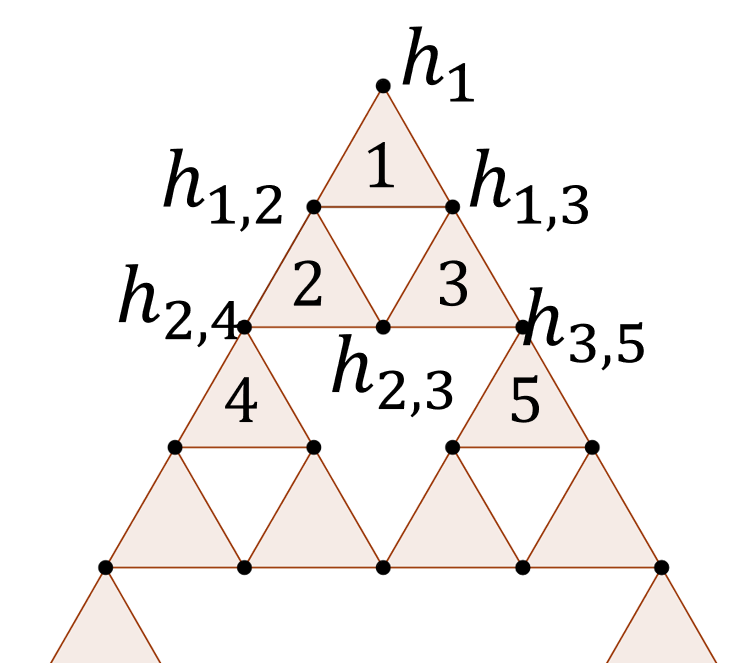}
 \end{center}
 \caption{Sierpinski gasket}
 \label{Sierpinski}
\end{figure}

The CG of this model is the Sierpinski gasket in Fig \ref{Sierpinski},
where the operators at vertices are given by
\begin{align}
    &h_1=\s_1 ^z, \nonumber\\
    &h_{1,2}=\s_1 ^x \s_2 ^z, \quad h_{1,3}=\s_1 ^y \s_3 ^z, \nonumber \\
    &h_{2,3}=\s_2 ^y \s_3 ^x, \quad h_{2,4}=\s_2 ^x \s_4 ^z, \quad
    h_{3,5}=\s_3 ^y \s_5 ^z, \nonumber \\
    &\cdots .
\end{align}
Since the Sierpinski gasket is a SPSC generated by 2-simplices, 
the Hamiltonian (\ref{eq:Hanoi}) can be transformed into a Majorana-bilinear form.
Note that the Sierpinski gasket is dual to the Hanoi graph.

\begin{figure}[htbp]
 \begin{center}
  \includegraphics[width=50mm]{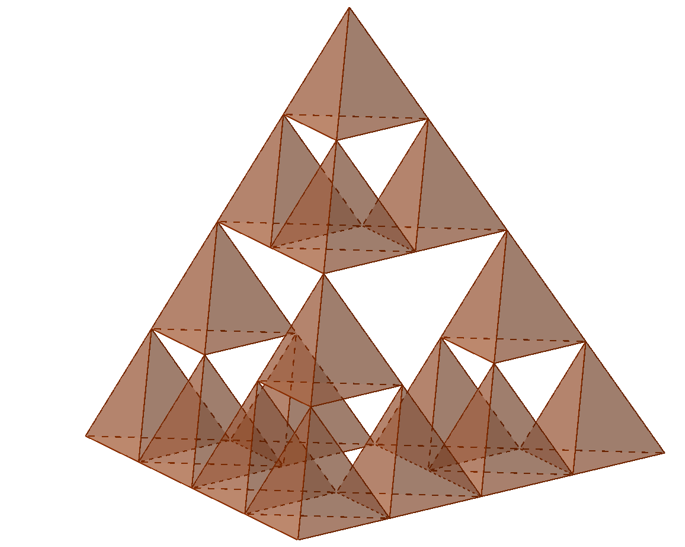}
 \end{center}
 \caption{Sierpinski tetrahedron}
 \label{Sierpinski tetrahedron}
\end{figure}


This model has 3d generalization.
Instead of the Hanoi graph, we use the dual lattice of the Sierpinski tetrahedron in Fig.\ref{Sierpinski tetrahedron}. Placing a Spin(4) generator at each site, we can construct the Hamiltonian of which the CG is the Sierpinski tetrahedron.
In the same way as the Hanoi graph, this model can be transformed into a Majorana-bilinear form.

\subsection{Octahedron model}
\begin{figure}[htbp]
 \begin{center}
  \includegraphics[width=70mm]{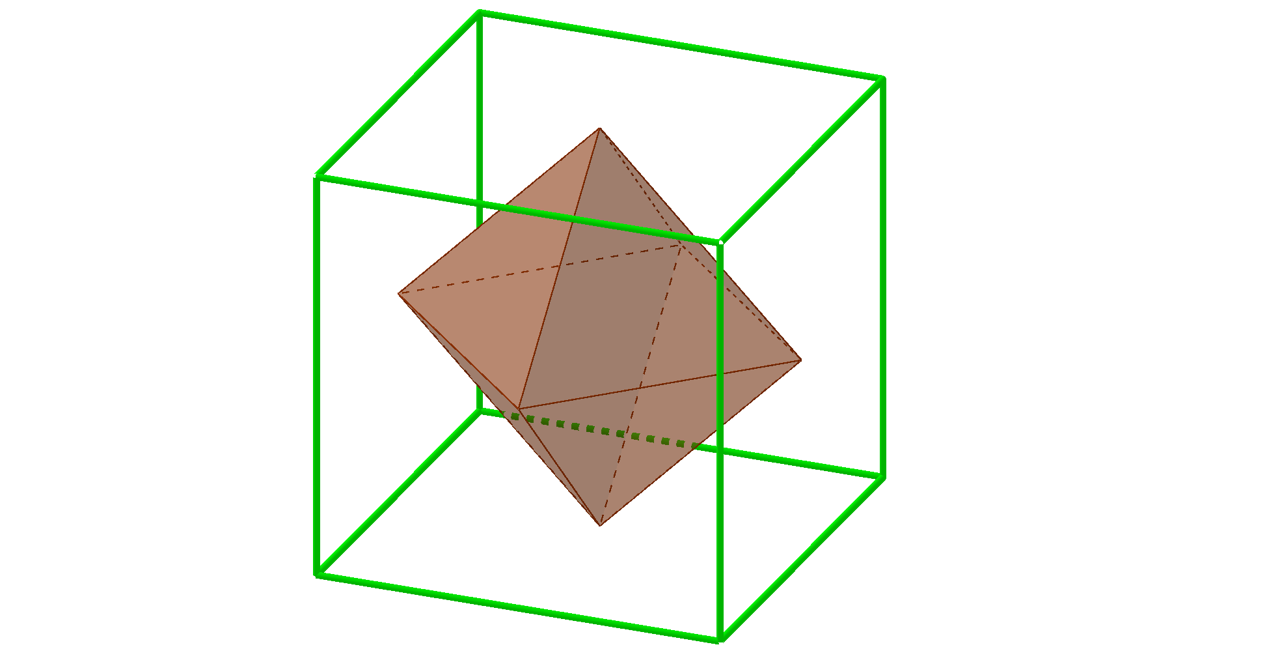}
 \end{center}
 \caption{Octahedron}
 \label{Octahedron}
\end{figure}

The dimension of simplices in a SPSC can be higher than the space dimension.
To illustrate this, we consider a spin model in the cubic lattice.
We place an SO(6) spin ({\it i.e.} a Spin(6) generator)
on each site of the cubic lattice, and consider the nearest neighbor interaction:
\begin{eqnarray}
H=-\frac{1}{2}\sum_{\bm{j}} \sum_{\mu=1} ^3 J_\mu \gamma^\mu_{\bm{j}} \gamma^{\mu+3}_{\bm{j}+\bm{e}_\mu} -g\sum_{\bm{j}}\gamma_{\bm{j}} ^7 ,
\end{eqnarray}
where $J_\mu$ is the exchange constant, $\gamma_{\bm j}^{\mu}$ is the SO(6) gamma matrix at the site ${\bm j}$, and 
${\bm e}_\mu$ is the unit vector in the $\mu$-th direction.
We assign operators
\begin{align}
    h_{\bm{j}} ^\mu=\gamma^\mu_{\bm{j}} \gamma^{\mu+3}_{\bm{j}+\bm{e}_\mu}, \quad
    h_{\bm{j}}'=\gamma_{\bm{j}} ^7.
\end{align}
The conserved quantities are 
\begin{align}
    c_{\bm{j}} ^{\mu,\nu}&=h_{\bm{j}} ^\nu h_{\bm{j}+\bm{e}_\nu} ^\mu h_{\bm{j}+\bm{e}_\mu} ^\nu h_{\bm{j}} ^\mu .
\end{align}
The CG of this model is vertex-sharing octahedra with central vertex in Fig.\ref{Octahedron}.
It is a SPSC since an octahedron with central vertex is a 6-simplex.
Thus, we can transform these operators into 
\begin{align}
    h_{\bm{j}} ^\mu=-i\epsilon^\mu_{\bm{j}} \p_{\bm{j}} \p_{\bm{j}+\bm{e}_\mu}, \quad
    h_{\bm{j}}'=-i\p_{\bm{j}} \p_{\bm{j}}' ,
\end{align}
and coserved quantities into
\begin{align}
c_{\bm{j}} ^{\mu,\nu}=\epsilon^\nu_{\bm{j}} \epsilon^\mu_{\bm{j}+\bm{e}_\nu} \epsilon^\nu_{\bm{j}+\bm{e}_\mu} \epsilon^\mu_{\bm{j}}.
\end{align}
Therefore, the Hamiltonian is recast into
\begin{eqnarray}
H=\frac{i}{2}\sum_{\bm{j}} \sum_{\mu=1}^3 J_\mu \epsilon^\mu_{\bm{j}} \p_{\bm{j}} \p_{\bm{j}+\bm{e}_\mu}+ig\sum_{\bm{j}}\p_{\bm{j}} \p_{\bm{j}}' .
\end{eqnarray}

In the following discussion, we take $g=0$ for simplicity.
In this case, $i\varphi'_{\bm j}\varphi'_{{\bm j}+{\bm e}_{\mu}}$ conserves, which induces additional $2^{N/2}$-fold degeneracy with the number of vertices $N$.
From Lieb's theorem \cite{lieb2004flux}, the ground state is realized when $\epsilon^\mu_{\bm{j}}=1$.
In this case, the Hamiltonian becomes
\begin{eqnarray}
H=\frac{i}{2} \sum_{\bm{j}} \sum_{\mu=1} ^3 J_\mu \p_{\bm{j}} \p_{\bm{j}+\bm{e}_\mu}.
\end{eqnarray}
By the Fourier transformation, 
\begin{eqnarray}
\p_{\bm{j}}=\int\frac{d^3 p}{(2\pi)^3} \left( e^{i\bm{p} \cdot \bm{j}} a_{\bm p}+ e^{-i\bm{p} \cdot \bm{j}} a_{\bm p} ^\dagger  \right), 
\end{eqnarray}
we have
\begin{align}
&H=\frac{i}{2}\sum_{\mu=1} ^3 J_d \int \frac{d^3 p}{(2\pi)^3} \left[e^{ip _\mu} a_{\bm{p}} a_{-\bm{p}}+e^{-ip_\mu} a_{\bm{p}} ^\dagger a_{-\bm{p}} ^\dagger
\right.
\nonumber\\
&
\left.
+e^{ip_\mu} a_{\bm{p}} ^\dagger a_{\bm{p}} +e^{-ip_\mu} a_{\bm{p}} a_{\bm{p}} 
^\dagger \right].
\end{align}
%
%
By diagonalizing this, the quasi-particle spectrum $\varepsilon_{\bm p}$ is obtained as
\begin{align}
\varepsilon_{\bm p}=\sum_{\mu=1}^{3}J_\mu \sin p_\mu, 
\end{align}
where the negative energy states are occupied in the ground state.

\section{Proofs}
\label{sec:proof}
Now we prove our main results, Theorems 1-3, in Sec.\ref{sec:main}.
To prove Theorem 1, we examine the basic properties of the CG.
Let us consider a transformation of the operators 
\begin{align}
\{\ldots h_{p}, \ldots, h_{q}, \ldots\}
\mapsto 
\{\ldots h_{p}, \ldots, h_{p}h_{q}, \ldots\}.    
\label{eq:htohh}
\end{align}
Corresponding to this transformation, 
the CG is modified as follows: 
\begin{itemize}
\item[i)] 
Draw new lines from $h_{p}h_{q}$ to all the $h_{k}$'s 
that satisfy $h_{p}h_{k}=-h_{k}h_{p}$. 

\item [ii)]
If there exist two lines from $h_{p}h_{q}$ to $h_{k}$, 
these lines should be eliminated 
and there remains no line between $h_{p}h_{q}$ and $h_{k}$. 
\end{itemize}
Here the rule ii) corresponds to the fact that 
when $h_{p}$ and $h_{q}$ anti-commutate with $h_{k}$, 
then the product $h_{p}h_{q}$ commutes with $h_{k}$. 

We represent the modification i) ii) in terms of the adjacency matrix on $\mathbb{F}_2$:
Let $M({\cal A})$ be the adjacency matrix of the CG ${\cal G(A)}$, i.e. 
\begin{eqnarray}
M({\cal A})_{ij}
=
\left\{
\begin{array}{cl}
0 & (h_{i}h_{j}=h_{j}h_{i}) \\
1 & (h_{i}h_{j}=-h_{j}h_{i}). 
\end{array}
\right.
\end{eqnarray}
$M({\cal A})$ is symmetric and its diagonal elements are all $0$. 
The multiplication of $h_{p}$ to $h_{q}$ 
corresponds to the row and column additions of $M({\cal A})$, i.e. 
the $q$-th row is replaced by the sum of $q$-th and $p$-th row, and the $q$-th column is replaced by the sum of $q$-th and $p$-th column. 
The row and column additions are given by
\begin{align}
M({\cal A}) \mapsto P^{[p,q]T} M({\cal A})P^{[p,q]},    
\end{align}
where $P^{[p,q]}$ is an elementary matrix with the $(i,j)$-component $P^{[p,q]}_{ij}=\delta_{ij}+\delta_{ip}\delta_{jq}$.
Here the rule $1+1=0$ in the matrix corresponds to the rule ii) above. 

We can also represent Eq.(\ref{eq:htohh}) using the same elementary matrix $P^{[p,q]}$:  
Let ${\bm v}(h_j)$ be the unit vector on $\mathbb{F}_2$ having a nonzero element only in the $j$-th component, 
\begin{align}
{\bm v}(h_j)=
\begin{pmatrix}
0 & \cdots & 0 &1&0\cdots& 0 \\
\end{pmatrix}^T.
\end{align}
Then, we have
\begin{align}
P^{[p,q]}{\bm v}(h_j)=
\left\{
\begin{array}{ll}
{\bm v}(h_p)+{\bm v}(h_q) & \mbox{for $j=q$}, \\
{\bm v}(h_j)     & \mbox{for $j\neq q$}, 
\end{array}
\right.
\label{eq:Pv}
\end{align}
which reproduces Eq.(\ref{eq:htohh})
by regarding the addition ${\bm v}(h_p)+{\bm v}(h_q)$ as the product $h_p h_q$.
%

Now consider the following operations on the CG:
If there are vertices $h_{i}$ and $h_{j}$ 
that are connected to each other with a line,  
then multiply $h_{i}$ to all the vertices $h_{k}$ 
that satisfy $h_{k}h_{j}=-h_{j}h_{k}$, 
and multiply $h_{j}$ to all the vertices $h_{k}$ 
that satisfy $h_{k}h_{i}=-h_{i}h_{k}$. 
Then there remains no line beginning from $h_{i}$ and $h_{j}$, 
except a line between $h_{i}$ and $h_{j}$.
As a result, we obtain a graph consisting only of $h_{i}$ and $h_{j}$, 
and a graph with other vertices. 
Repeating the same procedure for the latter graph, 
we inductively obtain graphs 
composed of only pairs and those with isolated vertices. 

This modification leads to Theorem 1:
After the modification of the CG,
$M({\cal A})$ is block diagonalized 
with $r/2$ number of blocks with the form 
$
\displaystyle
\left(
\begin{array}{cc}
0 & 1 \\
1 & 0 
\end{array}
\right) 
$
and $n-r$ number of blocks with $0$
\footnote{This fact itself is already known in the context of the matrix theory 
(see for example Theorem 8.10.1 in Ref.\cite{godsil}).}. ($r$ is even.)
Here $r/2$ is the number of the pairs and $n-r$ is the number of the isolated vertices in the above.
Since $r$ coincides with ${\rm rank\:}M({\cal A})$, 
the number of the pairs is unique.
When $h$ belongs to the kernel of $M({\cal A})$, 
it is evident that $h$ commutes with all the $h_{i}$'s, and hence $[H, h]=0$.
Conversely, assume that $h=h_{j_{1}}h_{j_{2}}\cdots h_{j_{k}}$ satisfies $[H, h]=0$. 
Then, we find $hh_{i}=\epsilon_{i} h_{i}h$, $\epsilon_{i}=+1$ or $-1$, for all $h_{i}$. 
If $h$ is a constant, $h$ generates an isolated vertex, and belongs to the kernel.
Otherwise from the condition $[H, h]=0$
and the independence of $h_j$s, 
it is easy to derive that $h$ commutes with all $h_{1}$,\ldots, $h_{n}$, 
and hence $h$ belongs to the kernel of $M({\cal A})$. 
Therefore, Theorem 1 holds.


By nothing that the 
$
\displaystyle
\left(
\begin{array}{cc}
0 & 1 \\
1 & 0 
\end{array}
\right) 
$ block and the 0 block correspond to the Clifford algebras $Cl_2$ and $Cl_1$, respectively,  
the above modification process also implies Proposition:
\medskip
\begin{itembox}[l]{Proposition}
Let ${\cal A}(X)$ be the BA generated from the set of independent operators $X$, 
and $M({\cal A}(X))$ be its adjacency matrix.  
Then, we find 
${\cal A}(X)\simeq (Cl_2)^{r/2}\otimes (Cl_1)^{n-r}$, 
and ${\cal A}(X)\simeq {\cal A}(X')$ if and only if ${\rm rank\:}M({\cal A}(X))={\rm rank\:}M({\cal A}(X'))$. 
\end{itembox}
\medskip

In particular, when ${\cal A}$ gives the complete graph with $n$ vertices, {\it i.e.} a graph in which all vertices are connected to each other, 
and when we separate a pair of operators in a manner similar to the above, it is easy to convince that the remaining graph with $n-2$ vertices becomes again a complete graph.  
Iterating this procedure, 
we finally obtain $n/2$ pairs when $n$ is even, 
and obtain $(n-1)/2$ pairs and an isolated vertex when $n$ is odd. 
The inverse of this modification is always possible.  
Since the complete graph with $n$ vertices represents the Clifford algebra with $n$ operators $Cl_n$, the rank of the adjacency matrix of the Clifford algebra with $n$ operators 
is $n$ when $n$ is even, and $n-1$ when $n$ is odd. 
This corresponds to the known fact 
$Cl_{2n} \simeq Cl_{2}^{\otimes n}$ and  
$Cl_{2n+1} \simeq Cl_{2}^{\otimes n}\otimes Cl_{1}$. 
Therefore, Proposition implies that a BA ${\cal A}$ with $n$ operators coincides with the Clifford algebra if 
${\rm rank} M({\cal A})=n$ (${\rm rank} M({\cal A})=n-1$) 
for even (odd) n.    


Theorem 2 follows from the fact that  $h_j$ in Eq.(\ref{eq:hpp}) reproduces the BA of the CG that coincides with a SPSC:
Let $K(S)$ with $S=\{s_1,\dots, s_m\}$ be the SPSC for the BA, and assign a Majorana operator $\varphi_{\alpha}$ on each simplex $s_{\alpha}\in S$. 
As we mentioned in Remark (i) in Sec.\ref{sec:main}, without loss of generality, we can assume that any vertex $v$ of $s_{\alpha}\in S$ is shared by another $s_{\beta}\in S$ ($\beta\neq \alpha$). 
Moreover, only the two simplices share $v$ 
since $S$ is single-point-connected.
Under this assumption, we consider $h_j^0\equiv-i\epsilon_{\alpha\beta}\varphi_{\alpha}\varphi_\beta$ for the vertex $v_j$ with $h_j$, where $\varphi_{\alpha}$ and $\varphi_\beta$ are located on the simplices that share $v_j$.
Then, we find that $\{h^0_i,h^0_j\}=0$ ($[h^0_i,h_j^0]=0$) if $v_i$ and $v_j$ are (not) vertices of the same simplex.
These relations reproduce the BA of the SPSC, and thus, we can identify $h_j^0$ with $h_j$.

Finally, we prove Theorem 3.
For preparation, we first show the following Lemma:
\medskip
\begin{itembox}[l]{Lemma}
Let $K(S)$ with $S=\{s_1,\dots,s_m\}$ be a SPSC. Then we have 
\begin{align}
C_q(K(S))=C_q(K(s_1))\oplus\cdots\oplus C_q(K(s_m))\quad (q\ge 1),    
\label{eq:CCC}
\end{align}
where $C_q$ is the $q$-chain on $\mathbb{F}_2$, and $\oplus$ is the direct sum ({\it i.e.} $C_q(K(s_{\alpha}))\cap C_q(K(s_\beta))=\{0\}$ for $\alpha\neq \beta$).
We also have
\begin{align}
H_q(K(S))=0 \quad (q\ge 2).    
\label{eq:H2}
\end{align}
\end{itembox}
\medskip

The proof is as follow:
Since $K(S)$ consists of all faces of $s_1, \dots, s_m$, we have
\begin{align}
C_q(K(S))=C_q(K(s_1))+\cdots+ C_q(K(s_m))\quad (q\ge 1).        
\end{align}
Furthermore, it holds that  $C_q(K(s_\alpha))\cap C_q(K(s_\beta))= \{0\}$ for $\alpha\neq \beta$ and $q\ge 1$ since $K(S)$ is a SPSC.
Thus, Eq.(\ref{eq:CCC}) holds.
Equation (\ref{eq:H2}) immediately follows from Eq.(\ref{eq:CCC}): 
Since the boundary operator  $\partial$ maps a $q$-chain to $(q-1)$-chain as, 
\begin{align}
\partial: C_q(K(s_\alpha)) \to C_{q-1}(K(s_\alpha)), 
\end{align}
we obtain
\begin{align}
H_q(K(S))=H_q(K(s_1))\oplus\cdots\oplus H_q(K(s_m))\quad (q\ge 2), 
\end{align}
which turns to be zero because $H_q(K(s_\alpha))=0$ ($q\ge 1$).

Now we can show that $K(S)$ has $n-m+1$ independent non-contractible loops.
Let $h_j$ ($j=1,\dots, n$) be the generators of a BA and $S=\{s_1, \dots, s_m\}$ be a set of simplices of which $K(S)$
is a SPSC of the BA. Consider the Euler characteristic of $\chi(K(S))$, \begin{align}
&\chi(K(S))
\nonumber\\
&=\sum_{q=0}^{{\rm dim}K(S)}(-1)^q(\mbox{the number of $q$-faces in $K(S)$}),
\end{align}   
where a $q$-face is a $q$-simplex included in $K(S)$ (namely a $0$-face is a vertex of $K(S)$, a $1$-face is a hinge of $K(S)$, and so on.)
In terms of homology groups, $\chi(K(S))$ is also written as \cite{nakahara}
\begin{align}
\chi(K(S))=\sum_{q=0}^{{\rm dim}K(S)}(-1)^q{\rm dim}H_q(K(S)).
\end{align}
Since $K(S)$ is connected, we have
\begin{align}
{\rm dim}H_0(K(S))=1,        
\end{align}
and from Lemma, it holds that 
\begin{align}
{\rm dim}H_{q\ge2}(K(S))=0.    
\end{align}
Thus, ${\rm dim}H_1(K(S))$ is evaluated as
\begin{align}
&{\rm dim}H_1(K(S))
\nonumber\\
&=1-\chi(K(S))
\nonumber\\
&=1-\sum_{q=0}^{{\rm dim}K(S)}(-1)^q(\mbox{the number of $q$-faces in $K(S)$}).
\label{eq:dimH1}
\end{align}
We compare this with the Euler characteristic of $K(s_\alpha)$ defined by
\begin{align}
&\chi(K(s_\alpha))
\nonumber\\
&=\sum_{q=0}^{{\rm dim}s_{\alpha}}(-1)^q(\mbox{the number of $q$-faces in $s_\alpha$}).    
\label{eq:chisa}
\end{align}
As $s_\alpha$ is a simplex, we have
\begin{align}
\chi(K(s_\alpha))=1,     
\end{align}
and thus, summing the both sides of Eq.(\ref{eq:chisa}) for all $s_{\alpha}\in S$, we obtain
\begin{align}
m=\sum_{\alpha=1}^m\sum_{q=0}^{{\rm dim}s_{\alpha}}(-1)^q(\mbox{the number of $q$-faces in $s_{\alpha}$}).    
\label{eq:m}
\end{align}
On the other hand, as $K(S)$ is a SPSC, we have
\begin{align}
&\sum_{q=0}^{{\rm dim}K(S)}(-1)^q(\mbox{the number of $q$-faces in $K(S)$}) 
\nonumber\\
&=
\sum_{\alpha=1}^m\sum_{q=0}^{{\rm dim}s_{\alpha}}(-1)^q(\mbox{the number of $q$-faces in $s_{\alpha}$})-n
\label{eq:-n}
\end{align}
Combing Eqs.(\ref{eq:m}) and (\ref{eq:-n}) with Eq.(\ref{eq:dimH1}), we get
\begin{align}
{\rm dim}H_1(K(S))=n-m+1,    
\end{align}
which implies that there exist $n-m+1$ non-contractible loops in $K(S)$.

The $n-m+1$ non-contractible loops give $n-m+1$ conserved quantities: For each non-contractible loop, consider a product of $h_j$ on all vertices in the loop. Obviously, the product reduces to a constant if we rewrite it in terms of  Majorana fermions in Theorem 2. Thus, it conserves and Theorem 3 holds.


\vspace{3ex}
\noindent

\section{Discussion}
\label{sec:conclusion}
In this paper, we present a simple criterion for solvability of lattice spin systems on the basis of the graph theory and the simplicial homology. When the lattice systems obey a class of algebras with the graphical representations, the spin systems can be converted into free Majorana fermion systems. We illustrate the validity of our criterion in a variery of spin systems.

Our method may reveal interesting aspects of lattice spin systems.
After the conversion to Majorana bilinear forms, the lattice spin systems exhibit particle-hole symmetry, in a manner similar to superconductors, because of the self-conjugate property of Majorana fermions. 
Hence, they can be a kind of topological superconductors \cite{sato2017}, although the origin of particle-hole symmetry is completely different.
The Kitaev honeycomb lattice, for instance, exhibits a 2d non-abelian topological phase analogue to chiral superconductors, in the presence of time-reversal breaking perturbation \cite{kitaev2006anyons}.
Our approach provides a systematic way to explore other interesting topological superconducting phases in spin systems; 3d non-abelian topological phase \cite{sato2003, teokane2010}, gapless topological phases \cite{satofujimoto2010, baum2015, kobayashi2014, agterberg2017}, and topological crystalline superconductors \cite{shiozakisato2014, shiozakisatogomi2016}.
Searching such interesting phases is left for future work. 

\noindent
\section*{Acknowledgement}
This work was supported by a Grant-in-Aid for Scientific Research on Innovative Areas ``Topological Materials Science" (KAKENHI Grant No.~JP15H05855) from the Japan Society for the Promotion of Science (JSPS). This work was also supported by JST CREST Grant No.~JPMJCR19T2, Japan. M.S. was supported by KAKENHI Grant Nos.~JP17H02922 and JP20H00131 from the JSPS. K.M. was supported by JSPS KAKENHI Grant No. JP19K03668.

\smallskip
{\it Note added.\,---\,}After completion of this work, we became aware of a recent related work~\cite{chapman2020characterization}.

\bibliographystyle{apsrev4-1}
\bibliography{gcsl}

\end{document}